\documentclass[prd, nofootinbib, floatfix, notitlepage]{revtex4-1}

\usepackage{amsmath,amsfonts,amssymb}
\usepackage{mathrsfs}
\usepackage{graphicx}
\usepackage[english]{babel} 
\usepackage{hyperref} 
\usepackage[normalem]{ulem}
\usepackage{textcomp}
\usepackage{floatrow}

\usepackage[usenames,dvipsnames]{xcolor}
\usepackage{bbold}

\usepackage{hyperref}
\usepackage[normalem]{ulem}
\usepackage{textcomp}

\usepackage[utf8]{inputenc}
\usepackage{comment}
\usepackage{pgf}
\usepackage{physics}

\usepackage{aas_macros}

\usepackage[dvipsnames]{xcolor}

\begin{document}

\title{Testing the Quasar Hubble Diagram with LISA Standard Sirens}
\author{Lorenzo Speri${}^{1}$}
\email{Contact author: lorenzo.speri@aei.mpg.de}
\author{Nicola Tamanini${}^1$}
\author{Robert R.~Caldwell${}^2$}
\author{Jonathan R.~Gair${}^1$}
\author{Benjamin Wang${}^3$}

\affiliation{
${}^1$Max Planck Institute for Gravitational Physics (Albert Einstein Institute), Am M{\"u}hlenberg 1, Potsdam 14476, Germany}
\affiliation{${}^2$Department of Physics \& Astronomy, Dartmouth College, 6127 Wilder Laboratory, Hanover, New Hampshire 03755 USA}
\affiliation{${}^3$University of California - Los Angeles, Los Angeles, California 90095 USA
}
\date{\today}

\begin{abstract}

Quasars have recently been used as an absolute distance indicator, extending the Hubble diagram to high redshift to reveal a deviation from the expansion history predicted for the standard, $\Lambda$CDM cosmology. Here we show that the Laser Interferometer Space Antenna (LISA) will efficiently test this claim with standard sirens at high redshift, defined by the coincident gravitational wave (GW) and electromagnetic (EM) observations of the merger of massive black hole binaries (MBHBs). Assuming a fiducial $\Lambda$CDM cosmology for generating mock standard siren datasets, the evidence for the $\Lambda$CDM model with respect to an alternative model inferred from quasar data is investigated. By simulating many realizations of possible future LISA observations, we find that for $50\%$ of these realizations (median result) 4 MBHB standard siren measurements will suffice to strongly differentiate between the two models, while 14 standard sirens will yield a similar result in $95\%$ of the realizations. In addition, we investigate the measurement precision of cosmological parameters as a function of the number of observed LISA MBHB standard sirens, finding that 15 events will on average achieve a relative precision of 5\% for $H_0$, reducing to 3\% and 2\% with 25 and 40 events, respectively. Our investigation clearly highlights the potential of LISA as a cosmological probe able to accurately map the expansion of the universe at $z\gtrsim 2$, and as a tool to cross-check and cross-validate cosmological EM measurements with complementary GW observations.

\end{abstract}

\maketitle

\newcommand{\ls}[1]{\textcolor{red}{Lorenzo: #1} }


\section{Introduction}
\label{sec:intro}

A key goal for cosmology is to characterize the accelerated cosmic expansion \cite{Riess:1998cb,Perlmutter:1998np} and understand the underlying physics \cite{Caldwell:2009ix,Weinberg:2012es}. Major efforts are underway to measure the expansion history using a variety of techniques, across a wide span of redshift, in hopes of revealing a clue to the nature of dark energy. To date, the Hubble diagram has been constructed out to redshift $z \sim 1.4$ using the local distance ladder \cite{Riess:2016jrr} and type Ia supernovae as standard candles \cite{2012ApJ...746...85S}. In the near future, dedicated observatories and experiments such as DESI \cite{desi}, Euclid \cite{euclid}, the Rubin Observatory \cite{lsst}, and the Roman Space Telescope \cite{wfirst} will refine and extend the Hubble diagram out to $z \sim 3$. Other methods, using for example the cosmic microwave background (CMB) \cite{Aghanim:2018eyx}, can retrieve information on the cosmic expansion at higher redshift, but must assume a cosmological model in order to constrain the late-time expansion rate.

A new, direct method to measure the expansion history out to significantly higher redshift has recently been implemented, using quasars as absolute distance indicators \cite{Bisogni:2017dzv,Risaliti:2018reu}. Determination of the quasar luminosity distance relies on a phenomenological relationship between the ultraviolet and X-ray emissions of the accreting plasma \cite{2016ApJ...819..154L}. Risaliti \& Lusso \cite{Risaliti:2018reu} analyzed a large catalog of X-ray and UV observations of quasars to calibrate this relationship and build a Hubble diagram that extends out to $z \simeq 5.5$. They found that their quasar sample yields distances that are in agreement with the standard, $\Lambda$CDM cosmological model at $z \lesssim 1.4$. However, at larger redshifts there appears to be significant disagreement between the quasar distances and the $\Lambda$CDM model predictions \cite{Risaliti:2018reu,Lusso:2019akb}. This innovation necessarily raises new questions. Are quasars truly reliable distance indicators?

There are no clear answers at present, and secondary analyses of the quasar data have come to varying conclusions about the state of the standard cosmology \cite{Melia:2019nev,Khadka:2019njj,Yang:2019vgk,Velten:2019vwo}. Is there a crisis? After all, the local Hubble diagram has come under close scrutiny, recently, due to the apparent tension with parameter values inferred from the CMB \cite{Riess:2016jrr,Aghanim:2018eyx,Freedman:2017yms,Riess:2020sih}. For certain, there is a need for new data and new methods to probe the expansion history.

The nascent field of gravitational wave (GW) astronomy offers a way to address these and other questions, providing a new, independent probe of the expansion using standard sirens \cite{Schutz:1986gp,Schutz:2001re,Holz:2005df}. An absolute distance is obtained from the inspiral and merger gravitational waveform, which, coupled with an electromagnetic (EM) measurement of the redshift, yields a new, independent determination of the cosmological parameters, for example of the Hubble constant \cite{Abbott:2017xzu}. It is worth stressing that distance measurements with GWs are absolute, meaning that they do not depend on a specific cosmological model or other distance indicators.

The standard siren method has already been used to produce constraints on the Hubble constant $H_0$ with real data collected by the LIGO-Virgo interferometers \cite{Abbott:2017xzu,Fishbach:2018gjp,Soares-Santos:2019irc,Abbott:2019yzh}. In particular GW170817 \cite{TheLIGOScientific:2017qsa}, a binary neutron star merger, was the first multi-messenger event for which both GW and EM signals were observed. This first standard siren provided a measurement of $H_0 = 70.0{}^{+12.0}_{-8.0}\,{\rm km\, s^{-1}\, Mpc^{-1}}$~\cite{2017Natur.551...85A,2019PhRvX...9a1001A,LVCO2Cosmo}. Other GW events observed without EM counterparts can also be used as standard sirens. If the host galaxy cannot be identified, one can still cross-correlate the sky localization region of the GW event with a galaxy catalog and assign to each galaxy within this region a non-zero probability of being the host galaxy~\cite{Schutz:1986gp,DelPozzo:2011yh,Gray:2019ksv}. This ``statistical method'' has less constraining power per event than the counterpart method because the redshift is not known precisely. However, combining information from many GW events allows the statistical significance to be increased. Combining all the events observed in the first two LIGO and Virgo observing runs gave a result $H_0 = 68^{+14}_{-7}$ km s$^{-1}$ Mpc$^{-1}$~\cite{LVCO2Cosmo}.

Looking ahead, GW observations by LIGO/Virgo and proposed observatories, such as the Cosmic Explorer \cite{Reitze:2019iox} or the Einstein Telescope \cite{Sathyaprakash:2012jk}, will yield orders of magnitude more candidate standard sirens. Perhaps the boldest of these future plans is the Laser Interferometer Space Antenna (LISA) \cite{Audley:2017drz}, a GW observatory consisting of a fleet of three satellites in an Earth-trailing, heliocentric orbit, positioned at the vertices of a $2.5$ million km triangle. Using laser interferometry, LISA will measure GWs in the mHz frequency range. The loudest objects in this band of the GW sky are expected to be mergers of massive black hole binaries (MBHBs). Observations of the expected GW signal, and an EM counterpart by ground-based telescopes, offer the exciting possibility to extend the Hubble diagram out to $z \sim 7$ or higher \cite{Tamanini:2016zlh,Tamanini:2016uin}. This is the possibility we explore: the ability of deep-redshift standard sirens observed by LISA to validate or refute the expansion history inferred by the quasar Hubble diagram \cite{Risaliti:2018reu}. To do this, we generate mock LISA MBHB standard siren observations based on a flat $\Lambda$CDM cosmology and we adopt a Bayesian framework to test the deviation found in the quasar observations. We aim to estimate how many MBHB standard sirens are necessary to rule out a deviation of that size and to explore the power of LISA to constrain the matter density and Hubble parameter. Due to the general formulation of our analysis, this work can be also used to test with standard sirens other cosmological models against $\Lambda$CDM.

In the long lead time before LISA flies, it is clear that the quasar Hubble diagram will evolve. The X-ray satellite eROSITA \cite{Merloni:2012uf} will deliver approximately three million AGNs \cite{refId0}. On this basis, the Hubble diagram could be boosted by a sample of $\sim 10^{4}$ quasars, distributed up to $z\sim 3$, after various selection cuts and criteria are applied \cite{Lusso:2020fax}. Higher redshift quasars that stretch back to the epoch of reionization may be observed by Euclid \cite{euclid}, with estimates suggesting $50-200$ objects in the range $z=7-8$, and order $10$ at $z=8-9$ \cite{Barnett:2019rtg}. Follow-up observations of a few of these rare objects could yield luminosity distances reaching as far back as standard sirens. The number of objects in the catalog is not the limiting factor, however. The major barrier is the scatter in the X-ray and UV relationship and any underlying systematic effects. Time will tell if the quasar Hubble diagram can be refined with ensuing observations, and whether the discrepancy with the $\Lambda$CDM expansion history persists. In this paper, we take the tension at face value, as a target for standard sirens. Of course, LISA offers a unique view of the cosmos in its own right, and will provide an independent probe of the high-redshift expansion history.

This paper is organized as follows. In Sec.~\ref{sec:quasars} we introduce the quasar Hubble diagram constructed with quasar observations and we briefly summarise the main steps to reproduce the results found in Ref.~\cite{Risaliti:2018reu}. In Sec.~\ref{sec:LISAsirens} we describe how the mock LISA standard sirens observations are generated and we investigate how such observations can constrain the Hubble parameter and the matter density. We present our method of Bayesian hypothesis testing in Sec.~\ref{sec:BayesSec}, and our results, in terms of the number of LISA standard sirens needed to test the quasar Hubble diagram, in Sec.~\ref{sec:Results}. We conclude with a discussion of the results and assess future prospects in Sec.~\ref{sec:Discussion}. The code used for the analysis is available at \cite{code_repo}.

\section{Quasar Hubble Diagram}
\label{sec:quasars}


Cosmologists have long hoped that quasars, observed at redshifts $z \sim 6$ and greater, could provide a reliable distance indicator with which to chart the cosmic expansion history. These aspirations took a significant step forward when a phenomenological relationship between the quasar x-ray and ultraviolet luminosities
\begin{equation}
    \log_{10}L_X = \gamma \log_{10} L_{UV} + \beta,
    \label{eqn:Leqn}
\end{equation}
was obtained by Avni \& Tananbaum \cite{1982ApJ...262L..17A}. Here, $L_X$, $L_{UV}$ are the monochromatic luminosities at $2$~keV and $2500~\AA$. There is good physical reason to expect the UV and X-ray spectra are connected: the UV emission originates in the infalling accretion material onto the quasar, whereas the X-rays are emitted by the plasma of hot electrons surrounding the accretion disk that scatter off the UV photons. Measurement of the fluxes yields the luminosity distance, once the parameters $\gamma$, $\beta$ are determined.  In terms of the observed fluxes, Eq.~(\ref{eqn:Leqn}) becomes
\begin{equation}
\log_{10}F_X = \gamma \log_{10}F_{UV} + 2 (\gamma-1) \log_{10}H_0 D_L + \tilde\beta, 
\label{eqn:flux}
\end{equation}
where $\tilde\beta$ absorbs the introduction of $H_0$ into the log of the luminosity distance. One can then use flux and redshift measurements to simultaneously fit the parameters and a theoretical model $D_L$. A data set consisting of measured values of $\log_{10}F_X$, $\log_{10}F_{UV}$, $z$, and uncertainty $\delta(\log_{10}F_X)$ for $N=1598$ quasars has been collected by Risaliti \& Lusso \cite{Risaliti:2018reu}, and generously shared with us.

\begin{figure}[t]
\begin{center}
\resizebox{!}{6cm}{\includegraphics{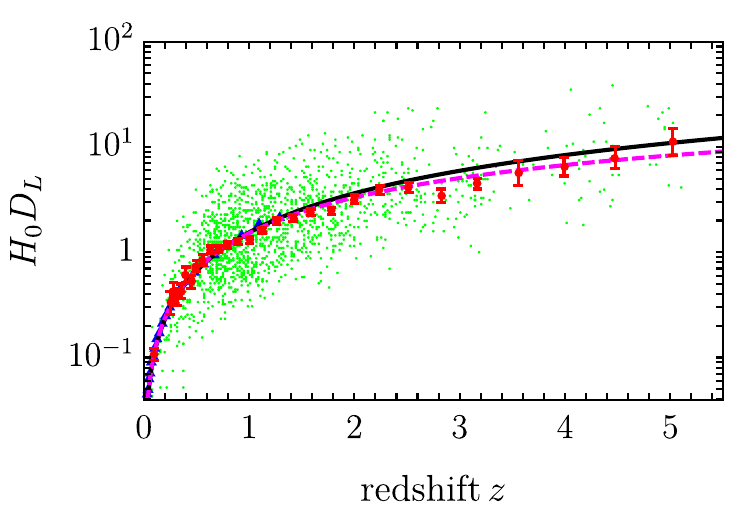}}
\hspace{0.4cm}
\resizebox{!}{6cm}{\includegraphics{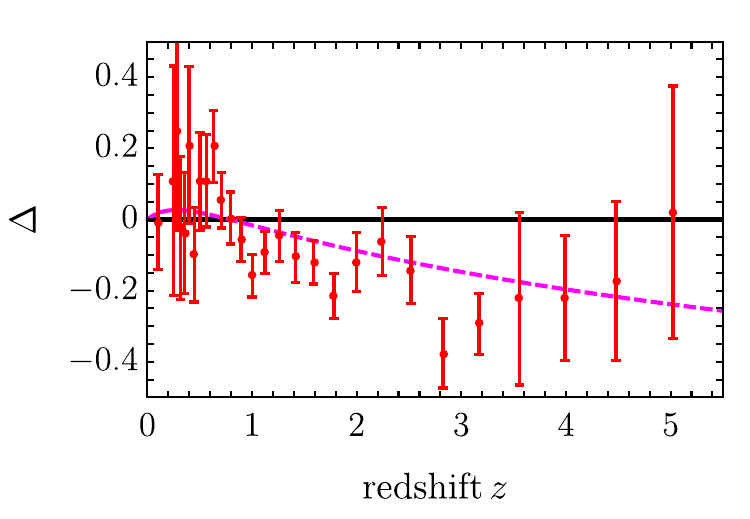}}
\caption{(Left) The quasar Hubble diagram based on the data set presented in Ref.~\cite{Risaliti:2018reu}, in the context of two cosmological models. Green dots are quasar data points, shown without error bars; red circles are the binned data, with errors. Blue triangles are the binned SN data, also shown without error bars. The solid black curve is the $\Omega_m=0.3$ $\Lambda$CDM model that best fits the joint quasar and type Ia supernovae at $z<1.4$. The dashed magenta curve is the best-fit phenomenological model proposed in Ref.~\cite{Risaliti:2018reu}. (Right) The relative difference between the luminosity distances for the binned quasar data relative to $\Lambda$CDM is shown as a function of redshift. The solid black line represents the $\Lambda$CDM baseline. The dashed magenta curve is the best-fit phenomenological model of Ref.~\cite{Risaliti:2018reu}.}
\label{fig:quasars}
\end{center}
\end{figure}

To constrain the luminosity distance-redshift relationship with quasar observations
we maximise the total likelihood of quasar data,
\begin{equation}
    {\cal L}_{QSO} = \prod_{i=1}^N \frac{e^{-\tfrac{1}{2}\Delta_i^2/\tilde\sigma_i^2}}{\sqrt{2 \pi \tilde\sigma_i^2}}.
    \label{eqn:Lqso}
\end{equation}
In the above equation, $\Delta_i = \log_{10}F_X|_i -( \gamma \log_{10}F_{UV} + 2 (\gamma-1) \log_{10}H_0 D_L + \tilde\beta)|_i$, $\tilde\sigma_i^2 = \delta^2 + \delta(\log_{10}F_X)|_i^2$, where $\log_{10}F_X|_i$ and $\delta(\log_{10}F_X)$  are respectively the observed log of the X-ray flux and its uncertainty, and $\delta$ introduces a new global variable to describe the intrinsic dispersion in the flux law, Eq.~(\ref{eqn:flux}). Following Ref.~\cite{Risaliti:2018reu}, we make a joint analysis with type Ia SNe data \cite{Betoule:2014frx}. The SNe likelihood is a multivariate gaussian distribution of the form  ${\cal L}_{SNe}\propto \exp(-\tfrac{1}{2}\chi^2_{SNe})$, where
\begin{equation}
    {\chi}^2_{SNe} = \sum_{ij}\Delta\mu_i C^{-1}_{ij} \Delta\mu_j,
    \label{eqn:Lsne}
\end{equation}
and $\Delta\mu = \mu_{obs} - \mu_{thy}$ is the difference in observed and theoretical distance modulus, and $C$ is the covariance matrix. The total likelihood is the product of ${\cal L}_{SNe}$ and ${\cal L}_{QSO}$. To begin, we use only those quasars in the same redshift range as the SNe ($z<1.4$). We bin the quasar data to match the redshifts of the SNe data set. In each bin, the mean log-flux is determined and its uncertainty is obtained. We maximize the sum of the log-likelihoods; the cosmological terms are common to both. In this way we calibrate the quasars, and thereby determine parameters $\gamma = 0.6,\, \tilde\beta=-14.7,\, \delta=0.25$; uncertainties in these parameters are subdominant to $\tilde\sigma$.  Given these parameters, the luminosity distance - redshift is fixed and Eq.~(\ref{eqn:flux}) yields the data set $\vec{y} = (z^i, \log _{10} (H_0 D_L ^i), \sigma \qty[\log_{10} (H_0 D_L  ^i)])$, shown in Fig.~\ref{fig:quasars}. For the redshift range used for calibration ($z<1.4$), we recover a $\Lambda$CDM model best fit value consistent with $\Omega_m = 0.31 \pm 0.05$ ($1\sigma$) as obtained in Ref.~\cite{Risaliti:2018reu}.

\begin{figure}[h]
    \centering
    \includegraphics{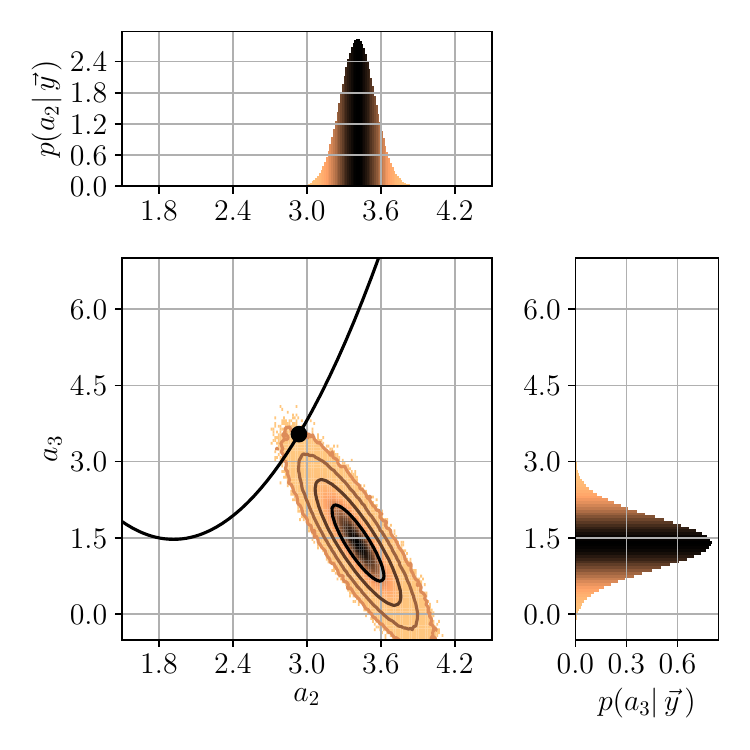}
    \caption{Posterior distribution of $a_2, a_3$ obtained with SNe and Quasar observations. The 1, 2, 3 and 4 $\sigma$ contours of the posterior distribution are shown as solid lines.
    A flat prior distribution was chosed for the MCMC on $(a_2,a_3)$.
    }
    \label{fig:a2a3_posterior}
\end{figure}{}

Proceeding to higher redshift, however, it is not immediately apparent whether the $\Lambda$CDM model and quasars are still in agreement. Fig.~\ref{fig:quasars} shows the binned quasar measurements depart from the $\Lambda$CDM prediction at high redshifts. To quantify this tension, Risaliti \& Lusso introduce an alternative (ALT) phenomenological model of the luminosity distance - redshift relation \cite{Risaliti:2018reu},
\begin{equation}
 D_L ^{\text{ALT} } (z; a_2, a_3) = \frac{c \, \naturallogarithm{10} }{H_0}\, ( x + a_2 x^2 + a_3 x^3 )\, , \qquad x=\logarithm _{10} (1+z)  \, .
\label{eqn:RL}
\end{equation}
The redshift-luminosity-distance relation $D_L ^{\text{ALT} }$ in Eq.~(\ref{eqn:RL}) is useful to encode the deviation found with the quasar observations.
Due to its general formulation, $D_L ^{\text{ALT} }$ can be also used to account for other kinds of observations or alternative cosmic expansion predictions which deviate from the $\Lambda$CDM. By expanding the luminosity distance of a flat $\Lambda$CDM cosmology at low redshift, it is possible to relate the coefficients $a_2, a_3$ of the phenomenological model in Eq.~\eqref{eqn:RL} with the matter density $\Omega _m$:
\begin{equation}
    a_2=3\ln(10)(2-\Omega_m)/4 \qquad \qquad
    a_3=\ln(10)^2(7/6 - 2 \Omega_m + 9 \Omega_m^2/8)
    \label{eqn:a2a3TOomega}
\end{equation}
Setting $\Omega_m=0.3$, we find that, with $a_2 \simeq 2.95$ and $a_3 \simeq 3.54$, $D_L ^{\text{ALT} }$ differs at most $0.6\%$ from the predictions of the flat $\Lambda$CDM for redshifts up to $z\sim 1$, and, at most $10\%$ for redshifts up to $z \sim 5$ \cite{banerjee2020cosmography}. 

Using the combined data set of SNe and quasars at all redshifts, it has been claimed \cite{Risaliti:2018reu} that the ALT model with $a_2 \simeq 3.5$ and $a_3 \simeq 1.5$ provides a much better fit to the data. Indeed, we reproduce the results of Ref.~\cite{Risaliti:2018reu} using the ensemble sampler for Markov Chain Monte Carlo (MCMC) \texttt{emcee} \cite{emcee}, and the data set $\vec y$ consisting of all 1598 quasars out to redshift $z \sim 5.5$ and the SNe observations to obtain the posterior distribution of $a_2, a_3$. The likelihood used for the MCMC is given by the product of a normal likelihood with mean given by the prediction of $D_L ^{\text{ALT} }$ for the quasar data and the aforementioned multivariate normal distribution ${\cal L}_{SNe}$ for the SNe data. The prior on $a_2, a_3$ is chosen to be a flat distribution. These choices allow us to recover the posterior distribution $p(a_i|\vec y)$ of $a_2, a_3$ shown in Fig.~\ref{fig:a2a3_posterior}, which is in good agreement with that of Figure 5 of Ref.~\cite{Risaliti:2018reu}.

It is possible to represent the flat $\Lambda$CDM model at low redshift in the $(a_2,a_3)-$parameter space by using the relation between the parameters $a_2, \,a_3$ and the matter density $\Omega _m$ in Eq.~\eqref{eqn:a2a3TOomega}. The black dot in Figure~\ref{fig:a2a3_posterior} corresponds to the specific value of $a_2, a_3$ for $\Omega _m = 0.3$, whereas the black solid line corresponds to other values of matter density. Although the relation between $a_2, a_3$ and the $\Lambda$CDM parameters is only approximate and based on an expansion at low redshift, Fig.~\ref{fig:a2a3_posterior} shows that the point corresponding to $\Lambda$CDM in the $(a_2,a_3)$ parameter space is $\sim4\sigma$ away from the best fit value in the $(a_2,a_3)$ posterior retrieved from quasar data. Based mainly on these results, it is further claimed \cite{Risaliti:2018reu} that $\Lambda$CDM is no longer a good fit to the quasar Hubble diagram at $z\gtrsim 3$, and, consequently, there is a $4\sigma$ tension between the standard cosmological model and the quasar data.

For the sake of argument, we can ask what does the phenomenological model have that $\Lambda$CDM does not? The arc of $H_0 D_L$ given by Eq.~(\ref{eqn:RL}) passes closer to the quasars at $z \sim 1.8,\, 3$ and above, than $\Lambda$CDM with $\Omega_m=0.3$. For visualization purposes, in Fig.~\ref{fig:quasars} we have binned the data to emphasize this point. In particular, the quasars at $2.7 \lesssim z \lesssim 3$ look like an outlier from the perspective of $\Lambda$CDM. This disagreement is made clearer where we plot the relative difference, $\Delta = (D_L^{QSO}/D_L^{LCDM})-1$, between the binned data and $\Lambda$CDM luminosity distance; the $\Lambda$CDM curve passes away from the binned data point by roughly four times its standard deviation. This difference is independent of binning in this redshift interval. However, we caution that the quasar data are sparse around both theoretical models at high redshift, suggesting that more data are needed. It is well to keep in mind that unresolved systematics, e.g. selection bias, quasar evolution, or accretion properties, may still affect the inferred luminosity-redshift relationship for quasars (see Ref.~\cite{2019ApJ...873L..12S}, but also Ref.~\cite{Salvestrini:2019thn}.)

Now we formulate our questions. We would like to find some complementary check of the Hubble diagram at $z \gtrsim 3$. Can we determine the luminosity distance with sufficient resolution to test deviations from $\Lambda$CDM at the 10\% level? Can we distinguish $\Lambda$CDM from the alternative illustrated in Fig.~\ref{fig:quasars}, at redshift $z \gtrsim 3$? These questions highlight a sort of new, high-redshift Hubble tension that can be addressed with GW standard sirens.

\section{LISA MBHB Standard Sirens}
\label{sec:LISAsirens}

LISA is a space mission designed to open the millihertz frequency range to GW observations \cite{Audley:2017drz}. In this frequency range different populations of GW sources are expected to be detected, including Galactic binaries \cite{Breivik:2017jip,Korol:2017qcx,Korol:2018ulo,Lau:2019wzw}, stellar-origin black hole binaries \cite{Sesana:2016ljz}, extreme mass ratio inspirals (EMRIs) \cite{Babak:2017tow}, the mergers of massive black hole binaries (MBHBs) with masses between $10^4$ to $10^7$ $M_\odot$ \cite{Klein:2015hvg} and perhaps stochastic GW backgrounds \cite{Caprini:2015zlo,Bartolo:2016ami,Caprini:2018mtu}. The population of MBHB sources is particularly interesting from a cosmological perspective, since MBHB mergers will be detected at high redshift (up to $z\sim 15-20$) and could have EM counterparts observable up to $z\sim 7$ \cite{Tamanini:2016zlh}.

As mentioned in Sec.~\ref{sec:intro}, obtaining complementary redshift information on GW events, through the identification of the corresponding host galaxy once an EM counterpart is spotted, is in fact one of the simplest ways to perform cosmological tests with GW observations. The GW events are standard sirens~\cite{Schutz:1986gp,Holz:2005df} that provide a direct measurement of the luminosity distance $D_L$ to the source. If this can be linked to a redshift $z$ obtained from EM observations, the combined observation provides a point on the distance-redshift relation $D_L(z;\vec{\theta} \, )$, which can be used to constrain the parameters, $\vec{\theta}$, of the cosmological model under consideration.

All GW sources that LISA will be able to observe at cosmological distances could be used as standard sirens, either with or without EM counterpart identification. For sources such as stellar-origin BHBs or EMRIs we do not expect any measurable EM counterpart (see however \cite{Eracleous:2019bal,Wang:2019bbk,Zhang:2019dpy}). For these sources, cosmological inference will only be possible using the statistical method~\cite{Kyutoku:2016zxn,DelPozzo:2017kme,MacLeod:2007jd,laghi2020}. MBHBs on the other hand could produce observable EM emissions to be detected at higher redshift with respect to stellar-mass BHBs and EMRIs. Several studies found that MBHBs could emit radiation in different bands of the EM spectrum both at merger and during long lasting (ranging from weeks to months) afterglows (see e.g.~\cite{Palenzuela:2010nf,2012AdAst2012E...3D,Giacomazzo:2012iv}). Moreover pre-merger EM observational signatures could even be spotted during their inspiral phase \cite{Kocsis:2007yu,OShaughnessy:2011nwl,2011ApJ...734L..37K,Haiman:2017szj}. Given a sufficiently accurate sky localization and a sufficiently powerful EM emission, future telescopes are expected to detect the counterpart of at least a few MBHB mergers per year \cite{Tamanini:2016uin,Tamanini:2016zlh}. These high-redshift standard sirens will allow us to probe the expansion of the universe at unprecedented distances and to test possible deviations from $\Lambda$CDM at $z\gtrsim3$ \cite{Caprini:2016qxs,Cai:2017yww,Belgacem:2019pkk}. This latter topic is the focus of our investigation, for which we will need realistic catalogs of LISA MBHB standard sirens. In what follows we will summarise how we construct such catalogs, building mainly on the results of \cite{Tamanini:2016uin,Tamanini:2016zlh}. The details of our approach are provided in Appendix~\ref{app:cats}.
 

\subsection{Construction of standard siren catalogs}
\label{subsec:SS_catalogs}


To simluate catalogs of MBHB standard sirens, we consider the source redshift distributions provided by \cite{Tamanini:2016zlh}, where the redshift distribution of LISA MBHB standard sirens with identified EM counterparts are reported according to the simulations performed in \cite{Tamanini:2016uin} and based on the LISA sensitivity curve of \cite{Audley:2017drz}. From these redshift distributions we pick suitable values for all events in the catalogs. To each redshift value we then associate a unique value of the luminosity distance $D_L$ by using the flat $\Lambda$CDM distance-redshift relation
\begin{equation}
    \label{eq:dist_flatLCDM}
            D_L ^{\Lambda\text{CDM}} (z; H_0,\Omega_{m}) = \frac{c \, (1 + z) \, }{H_0}\,  \int _0 ^ z \frac{d x}{\sqrt{\Omega_{m} (1+x)^3 + 1- \Omega_{m} }} \,,
\end{equation}
with fiducial cosmological parameters set to $H_0 = 70 \, {\rm km}\, {\rm s}^{-1}\, {\rm Mpc}^{-1}$ and $\Omega_m = 0.3$. The final value of the luminosity distance of each MBHB event in the catalog is then randomly scattered around the value given by our fiducial cosmology according to a Gaussian probability with standard deviation~$\sigma_{D}$, which is estimated as follows.

To each MBHB event in our simulated dataset we associate a realistic $1\sigma$ distance error by combining the following independent uncertainties (see Appendix~\ref{app:cats} for details):

\begin{itemize}
    \item The weak lensing contribution, estimated as in \cite{Tamanini:2016uin,Hirata:2010ba} but taking into account a possible delensing of up to 30\%.
    \item The peculiar velocity uncertainty contribution, estimated as in \cite{Tamanini:2016uin,Kocsis:2005vv}.
    \item The LISA instrument error, estimated to scale as $\sigma_{\rm LISA} / D_L \propto 2 D_L$ \cite{LiPhDThesis}.
    \item The redshift measurement error from the EM counterpart, assumed to be observed photometrically at $z>2$ and spectroscopically at $z<2$.
\end{itemize}

\begin{figure}
\begin{center}
\includegraphics[width=.7\textwidth]{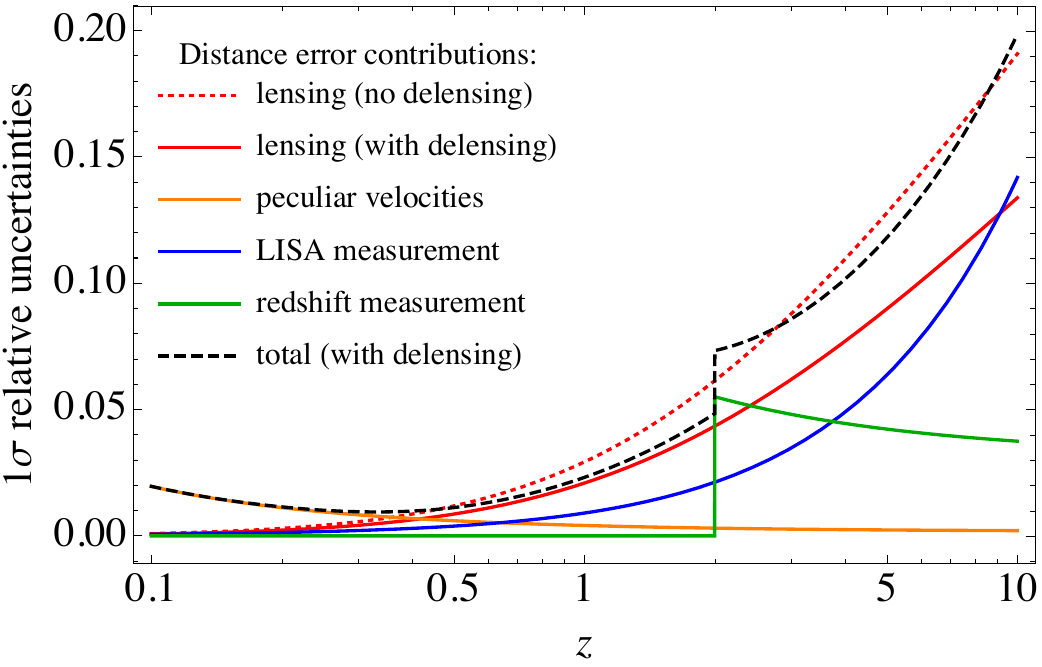}
\caption{
Relative uncertainty contributions to the luminosity distance of standard sirens ($\sigma_{D} / D_L$) as a function of redshift (see Sec.~\ref{subsec:SS_catalogs} and Appendix~\ref{app:cats} for details). For comparison, the average relative distance uncertainty of quasar data in \cite{Risaliti:2018reu} is 0.14.
}
\label{fig:errors}
\end{center}
\end{figure}

The total distance uncertainty $\sigma_{D}$ is then obtained by adding in quadrature all contributions listed above. Therefore, the total distance uncertainty depends not only on the source redshift but also on the assumed cosmological model. Fig.~\ref{fig:errors} shows the contribution of all these sources of error as a function of redshift, together with the total estimated error on the luminosity distance.

Each LISA MBHB event that will be used in the rest of the paper has been constructed according to the procedure outlined above and detailed in Appendix~\ref{app:cats}. Note however that this procedure cannot estimate the actual number of cosmologically useful events that LISA will detect. This information can only be obtained by simulating the full formation and evolution history of MBHBs, by performing parameter estimation over the GW signal and by estimating the emission and detectability of EM counterparts \cite{Tamanini:2016uin,NewMBHBCats}. Because of this all our results are given in terms of the number of standard siren that LISA will observe, with fiducial values taken from the existing literature.

\subsection{Testing $\Lambda$CDM with LISA MBHB standard sirens}

We can now use the catalogs defined above to check how well $\Lambda$CDM, the standard cosmological model, can be constrained by a given number of MBHB standard sirens. In particular we focus on how well the matter density parameter and the Hubble constant are constrained. For this reason, we express Eq.~(\ref{eq:dist_flatLCDM}) in terms of the Hubble parameter $h$ defined through $H_0/c = h/2997.9 \, \text{Mpc}^{-1}$, which for our fiducial flat $\Lambda$CDM is equal to $h = 0.7$. Since no more than $\sim$50 standard sirens are optimistically expected to be detected by LISA over its maximal extended mission duration (10 years) \cite{Tamanini:2016uin,Tamanini:2016zlh,Belgacem:2019pkk}, we will analyse the constraints arising from different possible observational outcomes, with the number of total detected standard sirens ranging from a minimum of 10 to a maximum of 50 events. Our fiducial scenario will assume 15 standard sirens, in agreement with a realistically expected outcome over a nominal mission of 4 years \cite{Tamanini:2016zlh,Belgacem:2019pkk}.

Given a set of standard sirens observations with associated luminosity distance errors, namely $\vec{y} = (z_j, D_j, \sigma _{D_j})$, we choose to work with the following likelihood,
\begin{equation}
\label{likelihood_distance}
    p(\vec{y}\,|\vec{\theta}, M_A ) = \prod_{j=1} ^{N_{SS}} \exponential \qty(- \frac{\qty(D_j - D^{M_A} _L (z_j; \vec{\theta}_A) )^2 }{2 \, \sigma_{D_j} ^2}) \, \frac{1}{\sqrt{2 \pi \sigma_{D_j} ^2}} \,,
\end{equation}
to conduct our statistical analysis. This specific choice implicitly assumes that the luminosity distance measurements are normally distributed around the cosmological predictions $D^{M_A} _L (z_j; \vec{\theta}_A)$ of a given cosmological model $M_A$ with parameters $\vec{\theta}_A$. In addition, the standard deviation of the normal likelihood is assumed to be the luminosity distance uncertainty $\sigma _{D_j} = \sigma _D (z_j)$ of the respective measurement, which in turn depends on the redshift and on the luminosity distance prediction of the cosmological model used to generate the data, i.e.~$\Lambda$CDM. In the following inference analysis we study the constraining power on the cosmological parameters of the $\Lambda$CDM model through the luminosity distance relation $D^{M_A} _L (z_j; \vec{\theta}_A)= D_L ^{\Lambda\text{CDM}} (z; h,\Omega_{m}) $. Therefore, we do not include the dependency of the luminosity distance uncertainty $\sigma _{D_j}$ on the cosmological model in our inference analysis and we take each $\sigma _{D_j}$ as an estimate for the measurement uncertainty (in other words the $\Lambda$CDM parameters over which $\sigma_{D_j}$ depends will not be varied, but fixed to their fiducial values).


\begin{figure}
\begin{center}
\includegraphics[width=.7\textwidth]{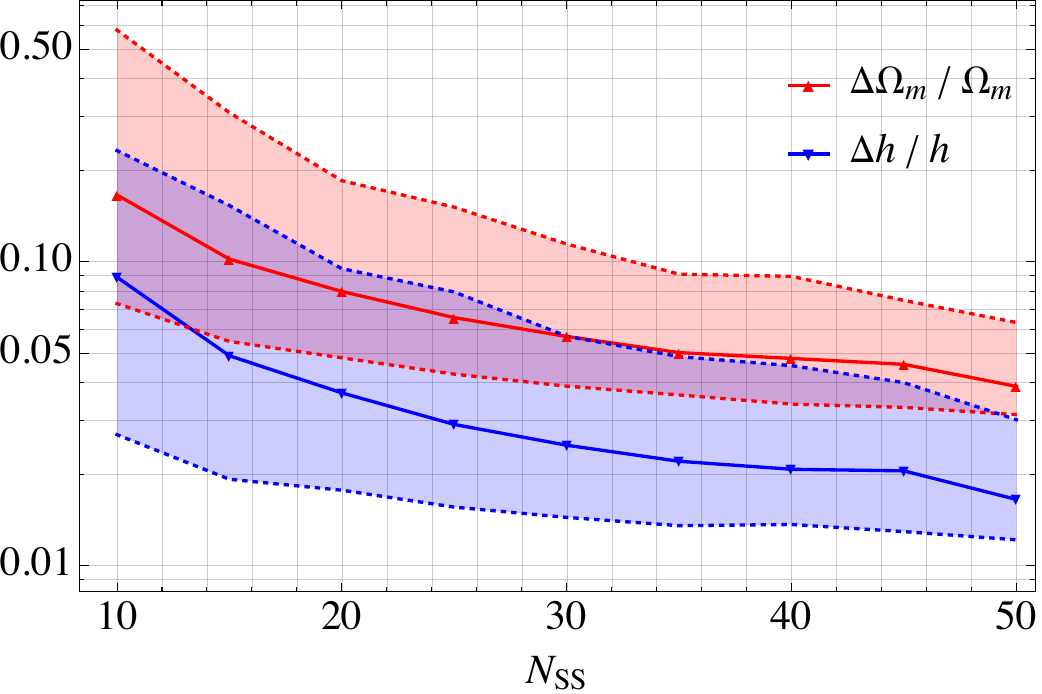}
\caption{
90\% C.I.~relative constraints on $\Omega_m$ and $h$ as a function of the number of standard sirens observed by LISA for 300 catalogue realizations. The triangles (and solid-line) indicate the median of the 90\% C.I.~relative constraint distribution of the 300 catalogs, whereas the shaded regions show the 90\% range of the 300 confidence intervals.
}
\label{fig:LCDM_constraints}
\end{center}
\end{figure}

In Fig.~\ref{fig:LCDM_constraints} we report the $\Lambda$CDM relative constraints at 90\% Confidence Interval (C.I.) as a function of the number $N_{SS}$ of standard siren events. To obtain these results, for each value of $N_{SS}$ we have simulated 300 standard siren catalogs. For each catalog, we find the best-fit $\Lambda$CDM cosmology through maximum likelihood estimation, which is equivalent to a least squares-fit to the $D_L$-$z$ measurements. Confidence intervals on the best-fit parameters are found by assuming Gaussian errors in the measurements, consistent with the model above. The numbers reported in Fig.~\ref{fig:LCDM_constraints} correspond to the median of the distribution of all the 90\% C.I.~obtained in this way from all the 300 catalogs with the same $N_{SS}$. In other words, the reported numbers represent the median of the corresponding quantity over 300 realizations with a fixed number of standard siren events, but different redshift and distance values for each event. The shaded regions show the central 90\% range of the 300 confidence intervals, for each choice of $N_{SS}$. From Fig.~\ref{fig:LCDM_constraints} it is clear that roughly 25 MBHB standard sirens will suffice to provide a $\sim$3\% error on $H_0$ and $\sim$6\% error on $\Omega_m$, while 40 MBHB standard sirens will improve these results to $\sim$2\% and $\sim$5\%, respectively. For our fiducial case however ($N_{SS}=15$) we can reach only a relative error of 5\% on $H_0$ and of 10\% on $\Omega_m$. These results are broadly in agreement with previous analyses \cite{Tamanini:2016uin,Belgacem:2019pkk}.


\begin{figure}
\begin{center}
\includegraphics[width=.5\textwidth]{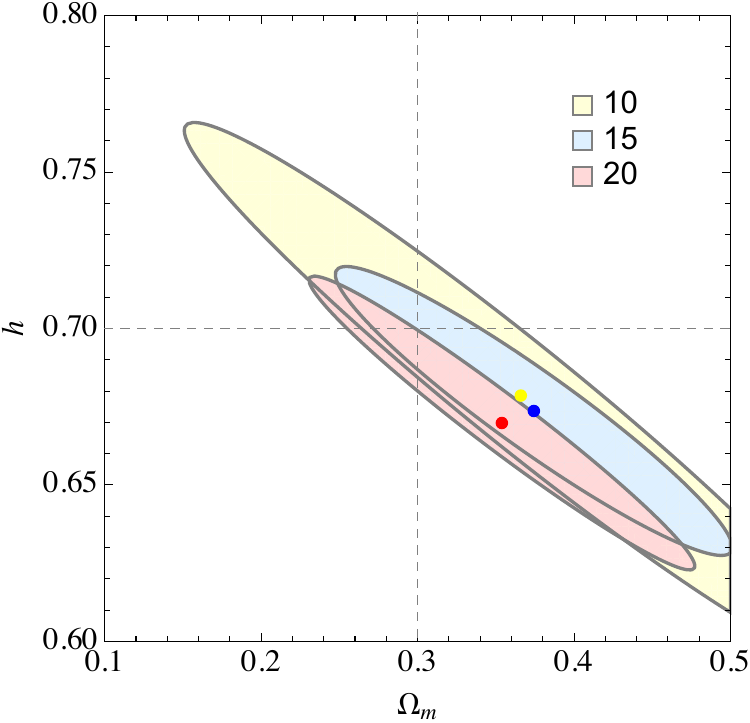}
\caption{
90\% C.I.~contour regions (Gaussian approximation) and best-fit values of the representative realizations of 10 (yellow), 15 (blue) and 20 (red) standard sirens, respectively, in the $\Lambda$CDM parameter space.
}
\label{fig:contour_ellipses}
\end{center}
\end{figure}

The constraining power in the $\Lambda$CDM parameter space depends on the specific realization. We estimate the constraining power of the measurements by computing a Gaussian approximation to the likelihood, which is
\begin{equation}
    \Gamma_{ij} = 
    \sum_n^{N_{SS}} \frac{1}{\sigma^2_{D_n}} \frac{\partial D^{\Lambda{\rm CDM}}_L (z_n; \vec{\theta}) }{ \partial \theta_i}
    \frac{\partial D^{\Lambda{\rm CDM}}_L (z_n; \vec{\theta}) }{ \partial \theta_j} \,,
\end{equation}
where $\vec\theta = (h,\Omega_m)$. The determinant of the matrix $\Gamma_{ij}$ determines the volume of the localization region. For each $N_{SS}$, we found the realization that had the median localization volume among all those simulated and took this to be ``representative'' of that number of standard sirens. In Fig.~\ref{fig:contour_ellipses} we use these representative realizations to construct contour regions at the 90\% confidence level in the $\Omega_m$-$h$ parameter space, for $N_{SS} = 10, 15, 20$. The contour regions in Fig.~\ref{fig:contour_ellipses} are based on the Gaussian approximation of the likelihood, and thus well characterize the latter only for a relatively large number of observations, which in our case means $N_{SS}\gtrsim 15$. The contour regions (ellipses) are centered on the best fit value estimated from the representative realization through the maximum likelihood estimator. Note that since in each dataset the data points have been scattered randomly around our fiducial cosmology, the actual values of the best fit parameters are not significant, as the best fit parameters scatter around the injected values depending on which redshift and measurement uncertainty are randomly drawn in the data generation procedure (the fact that all three best fit values lie on the bottom-right with respect to the true injected value is just a coincidence).

\begin{figure}[h]
    \centering
    \includegraphics[width=0.55\textwidth]{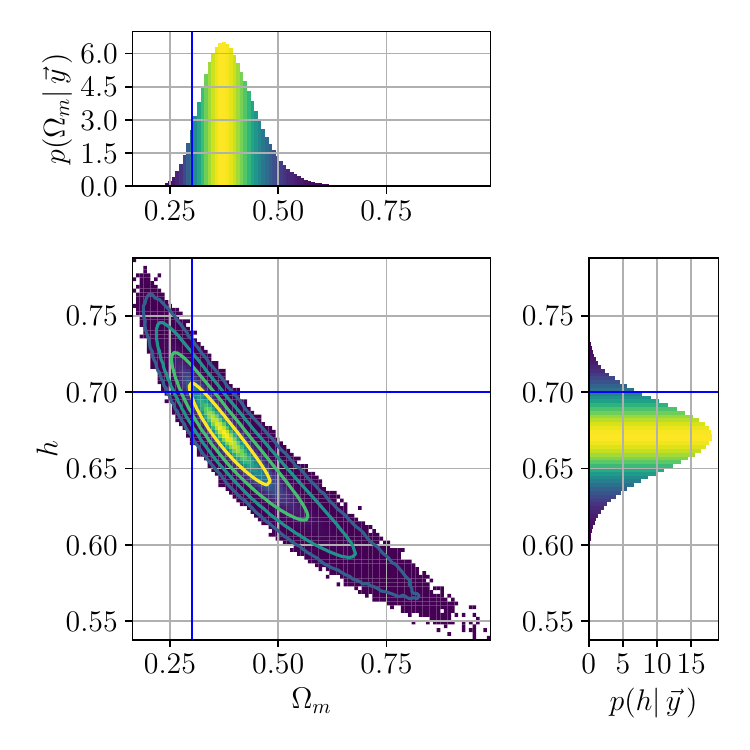}
    \caption{Posterior distribution of the Hubble parameter $h$ and matter density $\Omega_m$ obtained with the 15 representative standard siren observations $\vec{y}$. The median of the marginalized posteriors for matter density and Hubble parameter are respectively $0.384$ and $0.670$. The colored solid lines correspond to the $1,2,3,4\sigma$ contours, whereas the blue solid line shows the injected values $\Omega_m = 0.3$ and $h=0.7$ used to construct the data set. We used flat priors for both parameters $\Omega_m$ and $h$ with support $\Omega _m \in [0,1]$ and $h \in [0.2, 2.2]$, respectively.
    }
    \label{fig:posteriorLCDM}
\end{figure}

In order to further check the results for our fiducial scenario, and thus cross-check the Gaussian approximation, we have further performed a Bayesian analysis of the representative realization with 15 events. Therefore, we obtain the posterior distribution $p(\Omega_m, h|\vec{y})$ using \emph{emcee} \cite{emcee} and we show the results of the MCMC sampling in Fig.~\ref{fig:posteriorLCDM}. We also provide the MCMC implementation in Ref.~\cite{code_repo}. For the MCMC, we set flat priors for both parameters $\Omega_m$ and $h$ respectively with support $\Omega_m \in [0,1]$ and $h \in [0.2, 2.2]$ and we use the likelihood reported in Eq.~(\ref{likelihood_distance}).

\begin{table}[]
\begin{tabular}{|c|c|c|c|c|}
\hline
           & 68\% Credible Interval & 68\% Confidence Interval & Median & Best-Fit \\
           \hline
$\Omega_m$ &    $[0.327, 0.453]$    &$[0.313, 0.436]$&$0.384$&$0.374$\\
$h$        &$[0.648, 0.692]$&$[0.651, 0.696]$&$0.670$&$0.674$\\
\hline      
\end{tabular}
\caption{Summary statistics of the frequentist and Bayesian analysis of the 15 representative standard sirens shown in Fig.~\ref{fig:regression}}
\label{tab:fiducial_analysis}
\end{table}
{}
\begin{figure}[h]
\begin{center}
\resizebox{!}{6cm}{\includegraphics{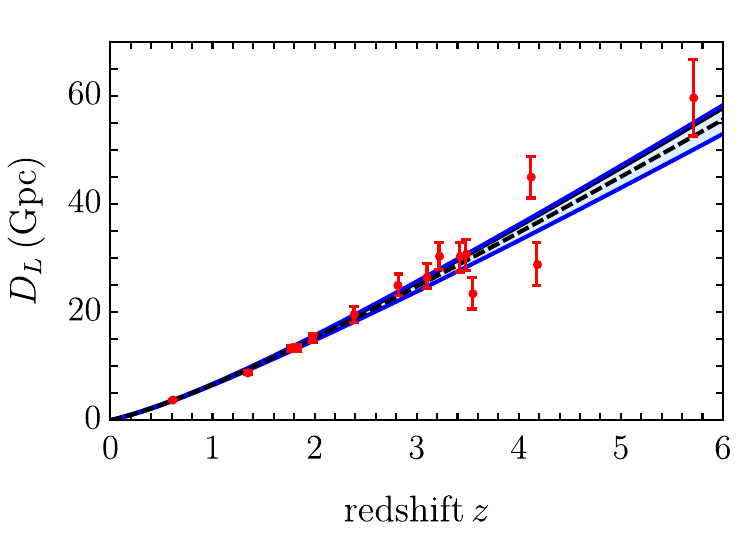}}
\hspace{0.4cm}
\resizebox{!}{6cm}{\includegraphics{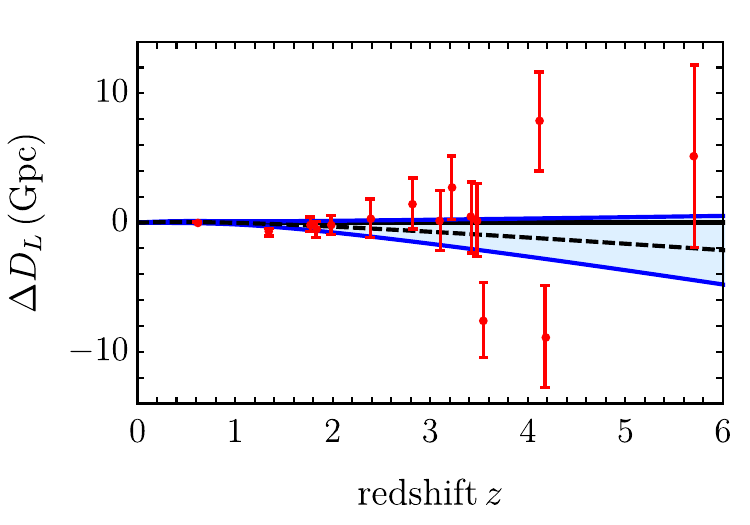}}
\caption{Representative LISA MBHB standard siren dataset with 15 events (red data points - 1$\sigma$ uncertainties reported). The left panel shows the distance-redshift space, while the right panel shows the relative difference with respect to our fiducial $\Lambda$CDM cosmology (black solid line). The black dashed line identifies our best fit to the dataset, while the solid blue lines delimit the 90\% C.I.~bands around it.}
\label{fig:regression}
\end{center}
\end{figure}

In Table~\ref{tab:fiducial_analysis} we show some summary statistics of the Bayesian and frequentist analysis of the 15 representative standard sirens. The median of the marginalized posterior shown in Fig.~\ref{fig:posteriorLCDM} are in agreement with the best-fit values obtained with the aforementioned frequentist analysis at $68 \%$  confidence level and both Bayesian and frequentist analysis yield $3\%$ and $16\%$ constraints on the relative precision of $h$ and $\Omega_m$, respectively.

Fig.~\ref{fig:contour_ellipses} shows that the higher the number of standard sirens, the better the $\Lambda$CDM constraints, as expected. By comparing Fig.~\ref{fig:contour_ellipses} and Fig.~\ref{fig:posteriorLCDM} we can also notice that the Gaussian approximation cannot fully catch all the features of the true posterior for 15 events, although the approximation seems reasonably correct at least to define 1 and 2$\sigma$ confidence regions.

Finally, in order to better visualize the uncertainty in the distance-redshift space as well, the representative realization with 15 standard sirens has been used to produce a regression plot in the $z$-$D_L$ space (left panel of Fig.~\ref{fig:regression}), which shows the allowed $\Lambda$CDM curves at the 90\% C.I.~(solid blue lines) together with the best fit from the dataset realization (dotted blue line), the injected $\Lambda$CDM curve (black solid line) and the data points of that particular realization with their associated 1$\sigma$ uncertainties in the distance (in red). The right panel of Fig.~\ref{fig:regression} shows the residuals of each curve and data point with respect to the fiducial $\Lambda$CDM curve. From Fig.~\ref{fig:regression} we deduce that LISA will test deviations from $\Lambda$CDM in the distance-redshift space at the 10\% level or better up to redshift $\sim$6, roughly.

This clearly shows the potential of LISA to provide unique cosmological constraints, not only for $\Lambda$CDM but also for other cosmological models, in a regime not yet efficiently probed by standard EM observations. These results already indicate that LISA will have the required power to test the deviations from $\Lambda$CDM reported with quasar data at $z\gtrsim 3$. In what follows we will quantify this statement using Bayesian hypothesis testing, and by looking in particular at how well LISA will differentiate between $\Lambda$CDM and the high-$z$ deviation found in Ref.~\cite{Risaliti:2018reu}.


\section{Bayesian cosmological hypothesis testing}

\subsection{Methodology}
\label{sec:BayesSec}

The scope of this section is to forecast the number $N_{SS}$ of MBHB standard sirens that we would need to observe to test the deviation from the $\Lambda$CDM model inferred from the quasar luminosity distances \cite{Risaliti:2018reu}. Bayesian hypothesis testing is a suitable framework for this goal, because the parameters of different cosmological models are not known precisely and are therefore well represented as distributions which depend on the model.

If we have two cosmological models $M_1$ and $M_2$ that describe the expansion history of the Universe and predict a luminosity distance relation $D_L ^{M_A} (z; \vec{\theta }_A)$, with parameters of the model $\vec{\theta }_A$, we can compute the Bayes factor 
\begin{equation}
    O_{12} = \frac{ p(\vec{y}\, | M_1 ) }{p(\vec{y}\, | M_2 ) } 
    \qquad
    \qquad
    p(\vec{y}\,| M_A ) = \int  p(\vec{y}\,|\vec{\theta }_A, M_A ) \, p(\vec{\theta }_A\,| M_A) \, \dd \vec{\theta }_A
    \qquad \qquad A=1,2
\end{equation}
to understand which model is favored by a set $\vec{y}$ of observations, assuming that the two models are equally probable a priori. 

In our analysis, the Bayes factor is used to forecast how many measurements $\vec{y} = (z_j,D_j, \sigma_{D_j} )$ of MBHB standard sirens are needed to conclude that the data strongly favor $\Lambda$CDM with respect to the deviation found in Ref.~\cite{Risaliti:2018reu}, assuming the underlying model providing the data is indeed $\Lambda$CDM. As observations, we use catalogues of luminosity distances $\vec{y} = (z_j,D_j, \sigma_{D_j} )$ generated using the prescription explained in Sec.~\ref{sec:LISAsirens}. The likelihood $p(\vec{y}\,|\vec{\theta }_A, M_A )$ and the model parameter priors $ p(\vec{\theta }_A\,| M_A) $, are chosen according to the following prescriptions. The likelihood $p(\vec{y}\,|\vec{\theta }_A, M_A )$  is given by Eq.~(\ref{likelihood_distance}) and the expected value $D^{M_A} _L (z_j; \vec{\theta}_A)$ is given by the prediction of luminosity distance of the following two cosmological models:

\begin{itemize}
    \item Model $1$: the flat $\Lambda$CDM model predicts the luminosity distance relationship given in Eq.~\eqref{eq:dist_flatLCDM}. Therefore we set $D_L ^{M_1} (z; \theta _1) =  D_L ^{\Lambda\text{CDM}} (z; \Omega_{m})$ where we consider the single parameter of the model to be the matter density $\Omega _m$.
    \item Model $2$: the ALT best fit-model\footnote{
    This is not truly a cosmological model, but it can represent a class of possible cosmological models that agree with the observations of Ref.~\cite{Risaliti:2018reu}. 
    }
    predicts the luminosity distance relationship given in Eq.~\eqref{eqn:RL}.
    Thus, we set $ D_L ^{M_2} (z; \vec{\theta}_2 ) = D_L  ^{\text{ALT}} (z; a_2, a_3) $, where the parameters of the model are $a_2$ and $a_3$.
\end{itemize}
We fixed in both models $H_0 = 70\, \text{km/s Mpc}$ and $c= 2.9979\times 10^8 \text{m/s}$ as done in Ref.~\cite{Risaliti:2018reu}. We do not include the Hubble constant as an additional parameter to Model 2 because we aim to test exclusively the predictions of the Universe expansion history of the two cosmological models, and because it is a calibration constant for the ALT best-fit model. Even including the Hubble constant as an additional parameter to Model 2, we find that our results are not considerably affected.
 
For Model~$1$ (flat $\Lambda$CDM model) a suitable prior is the beta p.d.f.~$\Omega_m \sim \beta (a,b)$ because of its support $\Omega_m \in [0,1]$. Notwithstanding the tight constraints on the value of $\Omega _m$ from different cosmological measurements (see e.g.~\cite{Aghanim:2018eyx}), we choose a prior peaked on the value of matter density of $\Omega_m = 0.31 \pm 0.05$ found in Ref.~\cite{Risaliti:2018reu} at low redshifts, where good agreement is obtained between the $\Lambda$CDM model and the quasar data. Therefore, we choose $\beta(a =  22.65, \,  b =  50)$ as the prior on the $\Lambda$CDM model in order to have a median of $0.31 \pm 0.05$, where the uncertainty is the $68\%$ credible interval. For comparison, the analysis of the CMB power spectrum combined with lensing and BAO give the following constraint on the matter density $\Omega _m  = 0.315 \pm 0.007$ \cite{Aghanim:2018eyx}.

For Model~$2$ (ALT best-fit model) we approximate the posterior of $a_2, a_3$ of Fig.~\ref{fig:a2a3_posterior} (Fig.~5 in Ref.~\cite{Risaliti:2018reu}) with a Multivariate Normal distribution $(a_2, a_3) \sim \mathcal{N}(\vec{\mu}, \Sigma ) $:
\begin{equation}
\label{RL_prior}
       p(\vec{\theta}_2|M_2) = p(a_2, a_2| M_2) =  \exponential \qty (-\frac{1}{2} \qty(\vec{\mathcal{A}} - \vec{\mu}) ^{\text{T}} \, \Sigma ^{-1} \, \qty(\vec{\mathcal{A}} - \vec{\mu}) ) \frac{1}{2 \pi \sqrt{\det \Sigma }} \qquad \qquad \vec{\mathcal{A}}= \mqty (a_2 \\ a_3).
\end{equation}
with covariance matrix $\Sigma$ and expected value $\vec{\mu} $ estimated from the MCMC posterior samples $a_2, a_3$ of Fig.~\ref{fig:a2a3_posterior} (Fig.~5 in Ref.~\cite{Risaliti:2018reu}):
\begin{equation}
\label{media_sigma}
   \vec{\mu} = \mqty(3.41309063 \\ 1.40248073) \qquad \Sigma = \mqty(  0.01961402 & -0.06005197 \\
 -0.06005197 & 0.24775043 ).
\end{equation}
The goodness of this approximation is confirmed by the very low Kullback-Leibler divergence ($\approx 4\times 10^{-7}$) between the approximate multivariate Gaussian distribution and true posterior distribution of  $a_2, a_3$. We use this multivariate normal distribution as the prior for Model 2, rather than the posterior samples, since it allows us to evaluate the double integral of the evidence $p(\vec{y}|M_2)$ analytically, as shown in Appendix~\ref{app:Bayes}. Having specified all the tools needed to calculate the Bayes factor, we can use several MBHB standard siren catalogues to study how many observations are necessary to test the claimed deviation from $\Lambda$CDM.

\subsection{Results}
\label{sec:Results}

We aim to study the Bayes factor $O_{12}$ as we add more observations. However, the Bayes factor itself depends on the redshifts of the observations, and measurements with different redshifts can favor one model over the other. Therefore we decided to study the distribution of Bayes factors by generating $10^4$ realizations of catalogues of $N_{SS}$ standard sirens.

\begin{figure}
\includegraphics{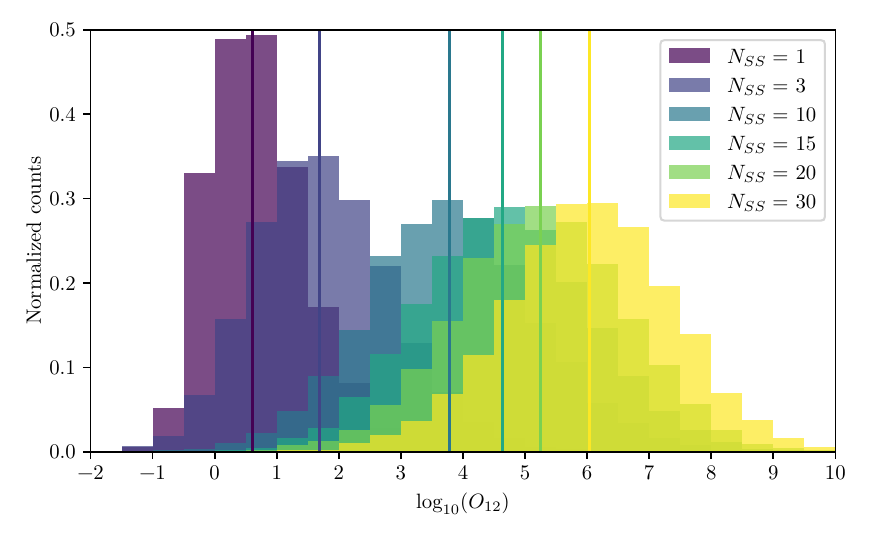}
\caption{Normalized distribution of $10^4$ Bayes factor $O_{12}$ realizations for given number $N_{SS}$ of standard siren observations. Here the Bayes factor encode how much the observations favour the $\Lambda$CDM model against the deviation observed with quasars. The solid vertical lines indicate the median of the distribution.}
\label{fig:bf_histogram}
\end{figure}

For each value of $N_{SS}$ we obtain the distribution of $O_{12}$ as follows
\begin{itemize}
    \item We draw a set of $N_{SS}$ redshifts from the expected distribution of MBHB standard sirens and assign uncertainties $\sigma_{D_j}$, as described in Sec.~\ref{sec:LISAsirens}.
    \item We scatter the luminosity distance measurements around the flat $\Lambda$CDM model with a Normal distribution $D_j \sim \mathcal{N}(D_L ^{\Lambda \text{CDM}}, \sigma_{D_j})$, as mentioned in Sec.~\ref{sec:LISAsirens}.
    \item We calculate the Bayes factor for the $N_{SS}$ simulated standard sirens $\vec{y} = (z_j,D_j, \sigma_{D_j} )$ using the likelihood in Eq.~(\ref{likelihood_distance}) and the priors of the two cosmological models with the aforementioned values of $a,\,b,\, \vec{\mu},\, \Sigma$.
\end{itemize}
We then obtain $10^4$ realizations of Bayes factors per each $N_{SS}$, which gives us a distribution of Bayes factors of possible future observations of MBHBs standard sirens. We can define the $\alpha \%$ \emph{Evidence Interval} $I = [O_L, O_R]$ as the symmetric interval of Bayes factors which covers the population of Bayes factor with probability of $\alpha \%$:
$$
\alpha \% = \int _{O_L} ^{O_R} p(O_{12}) \dd O_{12}. 
$$
Note that an evidence interval is neither a confidence interval nor a credible interval, because we are not estimating any parameters, but it tells us how likely we are to observe a realization of the Universe which allows us to distinguish between the two models. The distribution of the Bayes factors for the aforementioned priors is shown in Fig.~\ref{fig:bf_histogram}.

As we add more observations, the distribution of Bayes factors initially widens, but the median, represented by the vertical line, shifts to higher values. The shift in the median is due to the data favoring the $\Lambda$CDM model (Model~1) with respect to the ALT best-fit model (Model~2) as we increase the number of observations. This is of course expected since the standard siren data are generated according to the $\Lambda$CDM model.

The widening of the distribution is a consequence of the larger space of realizations of Bayes factor per $N_{SS}$.  In fact, if we draw a large number of $N_{SS}$ we can obtain different redshift realizations that can favor more or less one model over the other one. Those catalogues with many high-redshift standard sirens easily discriminate between the two models and have high Bayes factors, whereas catalogues with many low-redshift standard sirens have lower Bayes factors.
 

\begin{figure}
    \centering
    \includegraphics{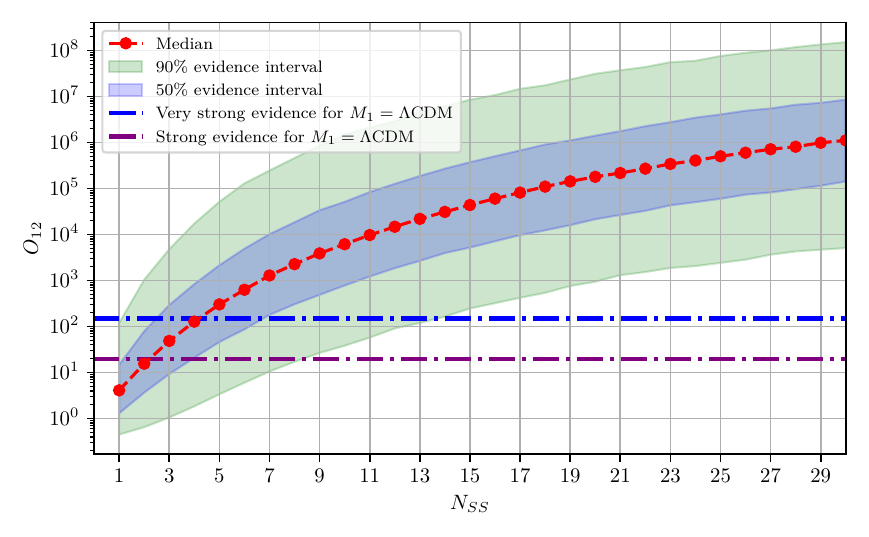}
    \caption{Bayes factor evolution as a function of the number of standard sirens $N_{SS}$. The red dots represent the median of the distribution of Bayes factor for given $N_{SS}$, and the shaded regions represent respectively the $90\%$ and $50\%$ evidence interval. The horizontal dashed lines represent the threshold above which the $\Lambda$CDM model is favored with respect to the ALT best-fit model, which encodes the deviation obtained from quasar observations. As a reference, we report that the Bayes factor of the 15 representative SS dataset of Fig.~\ref{fig:regression} is 6387.
    }
    \label{fig:BF_LCDM_RL}
\end{figure}

For Bayes factors above $\sim 20$ Model 1 ($\Lambda$CDM) is strongly favored with respect to Model 2, while for Bayes factors above $\sim 150$ the data very strongly favor Model 1 with respect to Model 2 \cite{kassBayesFactors1995}.
In order to assess how many standard sirens we need to distinguish between the two models, we plot in Fig.~\ref{fig:BF_LCDM_RL} the median, the $90\%$ and $50\%$ evidence interval of the Bayes factor distribution as a function of the number of standard sirens. In $50\%$ of realizations of the Universe, the Bayes factor will lie above the red line, under the aforementioned assumptions. From Fig.~\ref{fig:BF_LCDM_RL} it is clear that already with four standard sirens we have a $50\%$ chance that the Bayes factor would lie above the very strong evidence threshold, ruling out the deviation found in Ref.~\cite{Risaliti:2018reu}. With 14 MBHB standard sirens we have a $95\%$ chance that the two models will be distinguished by LISA.


\section{Discussion and conclusion}
\label{sec:Discussion}

Constructing the Hubble diagram is a fundamental step to study the expansion history of the Universe, but it requires accurate and precise measurements of the luminosity distance. The extension of the Hubble diagram with quasar observations has not only opened up the possibility to explore higher redshift regions but it has also suggested a possible new tension with the current cosmological model, the $\Lambda$CDM model. In this work we addressed the crucial question of whether and when we will be able to rule out or confirm this tension. We have every reason to expect the quasar Hubble catalog to grow and refine in the intervening years before LISA launches. Various estimates suggest a boost in the size of the quasar catalog by a factor of $10$ \cite{Lusso:2019akb} or even $100$ \cite{Risaliti_2015}. It is harder to forecast whether there will be a refinement in the intrinsic scatter of the quasar flux law. Hence, we have taken the quasar tension at face value in this work, as a target or reference against which to compare our forecasts of LISA standard siren measurements. Observations of GWs from MBHB mergers with associated EM counterparts will be critical to explore the high redshift regions of the Hubble diagram and, thus, to test the claimed deviation from $\Lambda$CDM.

Assuming the validity of the $\Lambda$CDM model, we described how to generate catalogs of mock MBHB standard siren observations by LISA.  We used a Bayesian approach to determine the relative favor of $\Lambda$CDM and the best fit model of Ref.~\cite{Risaliti:2018reu}, as determined by the data. This is expressed in terms of the Bayes factor, which can be calculated given priors of the model parameters, the likelihood function and the observed data. In the process, we reproduced some of the results of Ref.~\cite{Risaliti:2018reu}, in particular the posterior distribution of the phenomenological model, which we used as a basis for the prior beliefs on the deviation from $\Lambda$CDM. Given a fixed number of simulated standard sirens $N_{SS}$ we used $10^4$ catalog realizations to study the behavior of the Bayes factor and we found a 50\%, 75\% and 95\% chance to rule out the deviation found by Ref.~\cite{Risaliti:2018reu} with $N_{SS} = 4,7,14$, respectively.  These numbers are within the range of the expected rates for LISA MBHB standard sirens \cite{Tamanini:2016uin}, suggesting that LISA will indeed be able to discriminate the observed tension from the standard $\Lambda$CDM model. Furthermore, since the alternative hypothesis for the cosmic expansion has been formulated in a generic way, our analysis can also be used to test a wide variety of other cosmological models against the $\Lambda$CDM model with standard sirens. To facilitate this, our code for the Bayesian hypothesis testing analysis has been made available at \cite{code_repo}.
 
We have also evaluated the constraints on the matter density and Hubble constant as a function of the number of MBHB standard sirens and we found that the Hubble parameter can be more precisely constrained than the matter density on average. The relative precision of the Hubble parameter ranges from $3-20\%$ for 10 standard sirens up to $1-4\%$ for 50. Whereas, the matter density can be constrained with relative precision ranging from $7-50\%$ for 10 standard sirens up to $4-7\%$ for 50. The combination of MBHB events with other standard sirens detectable by LISA, in particular stellar-origin black hole binaries and EMRIs, will allow to improve the constraints on $H_0$ found in this paper, perhaps to below-percent accuracy in optimistic scenarios. These results will help elucidate the current tension on the measured value of the Hubble constant, which may well not be solved by the time LISA will launch.

The great potential of MBHB mergers for cosmology lies however beyond the results one can obtain on $H_0$. As we showed in this paper, by using these events as standard sirens LISA will probe the expansion history at $z\gtrsim 2$ with an accuracy possibly not attained by future EM observations, even as a new suite of large-scale structure surveys \cite{desi,euclid,lsst,wfirst,Bull:2020cpe} in the post-reionisation matter-dominated regime, $2 \lesssim z \lesssim 6$, are poised to search for clues to the nature of dark energy and chart the cosmic acceleration. Quasars have recently been proposed anew as a standard candles in this redshift range \cite{Risaliti:2018reu}. Other methods may also open up this frontier in the coming years, such as line intensity mapping \cite{Bernal:2019gfq,Karkare:2018sar,Kovetz:2017agg}. Reconciling the Hubble diagram among all these different methods will be a crucial activity in the future. LISA standard sirens offer a proven and complementary method to independently test the Hubble diagram at $z \gtrsim 2$. This synergy between GW and EM observations will provide the necessary confidence to reliably test the cosmic expansion at high redshift.


\acknowledgments
We thank Guido Risaliti and Elisabeta Lusso for kindly providing the quasar data from Ref.~\cite{Risaliti:2018reu}. We also thank Ollie Burke for his useful help and comments, Chiara Caprini for comments on the draft, and Ryan Hickox
for discussions on quasar astrophysics.


\appendix

\section{Details of the MBHB standard siren catalogs}
\label{app:cats}

In this appendix we provide the details of our approach to construct catalogs of MBHB standard sirens for LISA, as outlined in Sec.~\ref{subsec:SS_catalogs}. We start by drawing plausible redshift values for MBHB mergers from an interpolated redshift distribution based on the data presented in Fig.~1 of \cite{Tamanini:2016zlh}. These data are based on the LISA sensitivity curve adapted to the LISA design proposed in \cite{Audley:2017drz} and have been constructed using the procedure outlined in \cite{Tamanini:2016uin}, where MBHB merger histories have been reproduced using semi-analytical merger-tree simulations and the emission and detection of GW and EM signals have been estimated with standard astrophysical and data analysis techniques. We choose to work with the ``popIII'' model, which produces average results with respect to the astrophysical populations considered in \cite{Tamanini:2016uin,Tamanini:2016zlh}. In any case we checked that our final results do not change appreciably if the other astrophysical populations are considered, as the redshift distributions are similar (cf.~Fig.~1 of \cite{Tamanini:2016zlh}). We moreover set to zero the probability of drawing MBHB redshift values at $z<0.1$, in order to match our expectations of not finding MBHBs at very low redshift. We do not expect our results to be sensitive to this cut because the probability to have an event below $z = 0.1$ from our redshift distribution is particularly low, approximately $0.05\%$. To each redshift value in the catalog we then find a unique value of the luminosity distance $D_L$ by using the flat $\Lambda$CDM distance-redshift relation $D_L ^{\Lambda\text{CDM}} (z; H_0,\Omega_{m})$ as given in Eq.~\ref{eq:dist_flatLCDM}.
For ease of notation, we drop the label ${}^{\Lambda\text{CDM}}$ in this appendix.

To estimate a realistic $1\sigma$ distance error to each MBHB merger event, we combine the following uncertainties:
\begin{itemize}
    \item A weak lensing contribution analytically estimated by the following fitting formula \cite{Tamanini:2016uin,Hirata:2010ba}  (see also \cite{Cusin:2020ezb})
    \begin{equation}
        \frac{\sigma_{\rm lens}(z)}{D_L(z)} = 0.066 \left( \frac{1 - (1+z)^{-0.25}}{0.25} \right)^{1.8} \,.
    \end{equation}
We will consider the possibility of delensing, i.e.~the use of dedicated matter surveys along the line of sight of the GW event in order to estimate the lensing magnification distribution and thus remove part of the uncertainty due to weak lensing. Following \cite{Shapiro:2009sr} we will realistically assume that up to a 30\% delensing will be achievable at redshift 2 (note that in the most optimistic scenario this reduction could be as large as 50\%  \cite{Shapiro:2009sr}). Furthermore the delensing power should grow linearly from redshift zero, until it reaches a constant value at around $z=2$ (see Fig.~4 of \cite{Shapiro:2009sr}). For these reasons we will employ a delensing factor provided by the following phenomenological formula
    \begin{equation}
        F_{\rm delens}(z) = 1 - \frac{0.3}{\pi / 2} \arctan\left(\frac{z}{z_*}\right) \,,
    \end{equation}
    where $z_* = 0.073$ has been fixed requiring that $F_{\rm delens}(z)$ be within 1\% of 0.7 (the $z\rightarrow\infty$ asymptotic value) at $z=2$.
    The final lensing uncertainty on $D_L$ will thus be
    \begin{equation}
        \sigma_{\rm delens}(z) = F_{\rm delens}(z) \sigma_{\rm lens}(z) \,.
    \end{equation}
    \item A peculiar velocity uncertainty contribution as estimated by the following fitting formula \cite{Tamanini:2016uin,Kocsis:2005vv}
    \begin{equation}
        \frac{\sigma_v(z)}{D_L(z)} = \left[ 1 + \frac{c (1+z)^2}{H(z)D_L(z)} \right] \frac{\sqrt{\langle v^2 \rangle}}{c} \,,
    \end{equation}
    where the r.m.s.~peculiar velocity value is set to $\langle v^2 \rangle = 500$ km/s, in agreement with average values observed in galaxy catalogs.
    \item A LISA instrument error estimated as $\sigma_{\rm LISA} / D_L \propto 2 / {\rm SNR}$ \cite{LiPhDThesis}. If we further assume that, once marginalised over all other waveform parameters, the SNR scales as $1/D_L$, we can take the LISA measurement relative error on $D_L$ to be $\sigma_{\rm LISA} / D_L \propto 2 D_L$. To fix the constant of proportionality we rely on recent parameter estimation results \cite{Marsat:2020rtl}, which yield a 1$\sigma$ error on $D_L$ of roughly 5\% at redshift 4, by employing a full Bayesian approach over a MBHB signal with SNR = 945.
    The LISA measurement error on $D_L$ that we use is thus be given by
    \begin{equation}
        \frac{\sigma_{\rm LISA}(z)}{D_L(z)} = 0.05 \left(\frac{D_L(z)}{36.6\, {\rm Gpc}}\right) \,,
    \end{equation}
    where 36.6 Gpc corresponds to $z=4$ in our fiducial cosmology.
    \item A redshift error for all events at $z>2$ associated to photometric measurements estimated as \cite{Dahlen:2013fea,Ilbert:2013bf,Kruhler:2010jw}
    \begin{equation}
        \sigma_{\rm photo}(z) = 0.03 (1+z) \quad{\rm if}\; z>2 \,.
    \end{equation}
    For the sake of simplicity we propagate this redshift uncertainty to the distance uncertainty by assuming our fiducial cosmology, similarly to \cite{Tamanini:2016uin}.
    We assume redshift measurements at $z<2$ are taken spectroscopically and, for our purposes, with negligible errors (see e.g.~\cite{Dawson:2015wdb,Amendola:2016saw}).
\end{itemize}

Finally, as mentioned in the main text, the total distance uncertainty $\sigma_{D}$ is then obtained by adding in quadrature all contributions listed above (see Fig.~\ref{fig:errors}). Moreover the final value of the luminosity distance of each MBHB event is then randomly scattered around the value given by our fiducial cosmology, according to a Gaussian probability with standard deviation~$\sigma_{D}$.
This procedure delivers simulated of LISA MBHB standard sirens, similar for example to the one exposed in Fig.~\ref{fig:regression}.

\section{Details of the Bayesian methods}
\label{app:Bayes}
The evidence is the integral of the product between the likelihood and the model prior. Here we show how it is possible to calculate explicitly the evidence for the ALT best-fit model (Model 2) with the likelihood and model prior given respectively in Eqs.~(\ref{likelihood_distance}) and (\ref{RL_prior}).

Let us expand the argument of the exponential of Eq.~(\ref{likelihood_distance}):
\begin{equation}
        \ln\tilde{\cal L} =- \frac{\qty(D_j - D^{\text{ALT}} _L (z_j; a_2,a_3) )^2 }{2 \, \sigma_{D_j} ^2}
        =
        -
        \frac{\xi ^2}{2} \sum _j 
        \frac{(D_j/\xi - x_j - x_j ^2 a_2 -x_j ^3 a _3)^2}{\sigma^2 _{D_j}} \, ,
\end{equation}
where we defined $\xi = c \ln 10 /H_0$ and the sum over $j$ is the sum over the SS measurements expressed in terms of $x_j = \log _{10} (1+z_j)$. We now define $\psi_j= D_j/\xi - x_j $ and we express the above sum as
\begin{equation}
        \ln\tilde{\cal L}=
        -
        \frac{\xi ^2}{2}
        \sum _j \qty{
        \frac{\psi_j ^2}{\sigma^2 _{D_j}} + 
        a_2 ^2 \frac{x_j ^4}{\sigma^2 _{D_j}} + 
        a_3 ^2 \frac{x_j ^6}{\sigma^2 _{D_j}} 
        -2 a_2 \frac{\psi_j x_j ^2}{\sigma^2 _{D_j}} 
        -2 a_3 \frac{\psi_j x_j ^3}{\sigma^2 _{D_j}} 
        +2 a_2  \, a_3  \frac{ x_j ^5}{\sigma^2 _{D_j}}
        } \, .
\end{equation}
If we express the above equation in terms of:
\begin{equation}
    \vec{\mathcal{A}}= \mqty (a_2\\ a_3)
    \qquad \qquad 
\mathcal{C}= \mqty (
\sum _j \frac{x_j ^4}{\sigma^2 _{D_j}} &
\sum _j \frac{ x_j ^5}{\sigma^2 _{D_j}} \\ 
\sum _j \frac{ x_j ^5}{\sigma^2 _{D_j}} &
\sum _j \frac{x_j ^6}{\sigma^2 _{D_j}}
) \qquad \qquad 
\vec{\mathcal{B}}= \mqty (-2 \sum _j \frac{\psi_j x_j ^2}{\sigma^2 _{D_j}}  \\
-2 \sum _j \frac{\psi_j x_j ^3}{\sigma^2 _{D_j}} 
)
\end{equation}
we obtain that the likelihood Eq.~(\ref{likelihood_distance}) can be expressed as:
\begin{equation}
p(\vec{y}\,|a_2, a_3)=
\exp \qty{-\frac{\xi ^2}{2} \qty [
\sum _j \frac{\psi_j ^2}{\sigma^2 _{D_j}} +
    \vec{\mathcal{A}} ^{\text{T}} \mathcal{C} \vec{\mathcal{A}} +
    \vec{\mathcal{B}} ^{\text{T}} \vec{\mathcal{A}}
] }  
\prod _j \frac{1}{\sqrt{2 \pi \sigma_{D_j} ^2}}
\end{equation}
In order to obtain the evidence for the ALT model we need to multiply the likelihood by the model prior in Eq.~(\ref{RL_prior}).
So the evidence for the ALT model is
\begin{equation}
\begin{split}
    p(\vec{y}| M_2)=&\frac{1}{2 \pi \sqrt{\det \Sigma }}
    \,
    \qty(\prod _j \frac{1}{\sqrt{2 \pi \sigma_{D_j} ^2}})
    \,
    \exp \qty(
    -\frac{1}{2}  \qty(\vec{\mu }) ^{\text{T}} \, \Sigma ^{-1} \, \vec{\mu}
    -\frac{\xi ^2}{2} 
\sum _j \frac{\psi_j ^2}{\sigma^2 _{D_j}}
    ) \times \\
    & \qquad \int 
    \exp \qty {
        \qty(-\frac{\xi ^2}{2} \vec{\mathcal{B}} ^{\text{T}} +  \qty(\vec{\mu }) ^{\text{T}} \, \Sigma ^{-1}) \vec{\mathcal{A}}
    -\frac{1}{2} 
    \qty(\vec{\mathcal{A}}) ^{\text{T}} \, \qty (
        \Sigma ^{-1} +
        \xi^2 \mathcal{C}
        ) 
    \vec{\mathcal{A}}
    }
    \, \dd a_2 a_3 = 
    \\ 
     =&\frac{1}{2 \pi \sqrt{\det \Sigma }}
    \,
    \qty(\prod _j \frac{1}{\sqrt{2 \pi \sigma_{D_j} ^2}})
    \,
    \exp \qty(
    -\frac{1}{2}  \qty(\vec{\mu }) ^{\text{T}} \, \Sigma ^{-1} \, \vec{\mu}
    -\frac{\xi ^2}{2} 
\sum _j \frac{\psi_j ^2}{\sigma^2 _{D_j}}) \times
    \\
    \qquad &
    \exp \qty(
    \frac{1}{2} \vec{\mathcal{V}} ^{\text{T}}
    \,
    \qty [
        \Sigma ^{-1} +
        \xi^2 \mathcal{C}
        ]^{-1}
        \,
    \vec{\mathcal{V}}
    ) 
    \frac{2 \pi}{\sqrt{\det \qty (
        \Sigma ^{-1} +
        \xi^2 \mathcal{C}
        )  }}
\end{split}
\end{equation}
where we introduced $\vec{\mathcal{V}} = -\frac{\xi ^2}{2} \vec{\mathcal{B}} +   \, \Sigma ^{-1} \vec{\mu }$ in the last line. The above equation shows how it is possible to directly evaluate the evidence of the ALT best fit model using the observations $\vec{y}$.

\bibliography{paper.bib}

\begin{thebibliography}{10}

\bibitem{Riess:1998cb}
A.~G. Riess et~al.
\newblock {Observational evidence from supernovae for an accelerating universe
  and a cosmological constant}.
\newblock {\em Astron. J.}, 116:1009--1038, 1998.

\bibitem{Perlmutter:1998np}
S.~Perlmutter et~al.
\newblock {Measurements of Omega and Lambda from 42 high redshift supernovae}.
\newblock {\em Astrophys. J.}, 517:565--586, 1999.

\bibitem{Caldwell:2009ix}
R.~R. Caldwell and M.~Kamionkowski.
\newblock {The Physics of Cosmic Acceleration}.
\newblock {\em Ann. Rev. Nucl. Part. Sci.}, 59:397--429, 2009.

\bibitem{Weinberg:2012es}
D.~H. Weinberg et al.
\newblock {Observational Probes of Cosmic Acceleration}.
\newblock {\em Phys. Rept.}, 530:87--255, 2013.

\bibitem{Riess:2016jrr}
A.~G. Riess et~al.
\newblock {A 2.4\% Determination of the Local Value of the Hubble Constant}.
\newblock {\em Astrophys. J.}, 826(1):56, 2016.

\bibitem{2012ApJ...746...85S}
N.~{Suzuki} et al.
\newblock {The Hubble Space Telescope Cluster Supernova Survey. V. Improving
  the Dark-energy Constraints above $z > 1$ and Building an Early-type-hosted
  Supernova Sample}.
\newblock {\em \apj}, 746(1):85, February 2012.

\bibitem{desi}
\url{https://www.desi.lbl.gov/}.

\bibitem{euclid}
\url{https://euclid.jpl.nasa.gov/}.

\bibitem{lsst}
\url{https://www.lsst.org/}.

\bibitem{wfirst}
\url{https://wfirst.gsfc.nasa.gov/}.

\bibitem{Aghanim:2018eyx}
N.~Aghanim et~al.
\newblock {Planck 2018 results. VI. Cosmological parameters}.
\newblock {\em Astron. Astrophys.}, 641:A6, 2020.

\bibitem{Bisogni:2017dzv}
S.~Bisogni, G.~Risaliti, and E.~Lusso.
\newblock {A Hubble diagram for quasars}.
\newblock 2017.

\bibitem{Risaliti:2018reu}
G.~Risaliti and E.~Lusso.
\newblock {Cosmological constraints from the Hubble diagram of quasars at high
  redshifts}.
\newblock {\em Nat. Astron.}, 3(3):272--277, 2019.

\bibitem{2016ApJ...819..154L}
E.~{Lusso} and G.~{Risaliti}.
\newblock {The Tight Relation between X-Ray and Ultraviolet Luminosity of
  Quasars}.
\newblock {\em \apj}, 819(2):154, Mar 2016.

\bibitem{Lusso:2019akb}
E.~Lusso et al.
\newblock {Tension with the flat $\Lambda$CDM model from a high-redshift Hubble
  diagram of supernovae, quasars, and gamma-ray bursts}.
\newblock {\em Astron. Astrophys.}, 628:L4, 2019.

\bibitem{Melia:2019nev}
F.~Melia.
\newblock {Cosmological test using the Hubble diagram of high-z quasars}.
\newblock {\em Mon. Not. Roy. Astron. Soc.}, 489(1):517--523, 2019.

\bibitem{Khadka:2019njj}
N.~Khadka and B.~Ratra.
\newblock {Quasar X-ray and UV flux, baryon acoustic oscillation, and Hubble
  parameter measurement constraints on cosmological model parameters}.
\newblock {\em Mon. Not. Roy. Astron. Soc.}, 492(3):4456--4468, 2020.

\bibitem{Yang:2019vgk}
T.~Yang, A.~Banerjee, and E.~O. Colgáin.
\newblock {On cosmography and flat $\Lambda$CDM tensions at high redshift}.
\newblock 2019.

\bibitem{Velten:2019vwo}
H.~Velten and S.~Gomes.
\newblock {Is the Hubble diagram of quasars in tension with concordance
  cosmology?}
\newblock {\em Phys. Rev.}, D101(4):043502, 2020.

\bibitem{Freedman:2017yms}
W.~L. Freedman.
\newblock {Cosmology at a Crossroads}.
\newblock {\em Nat. Astron.}, 1:0121, 2017.

\bibitem{Riess:2020sih}
A.~G. Riess.
\newblock {The Expansion of the Universe is Faster than Expected}.
\newblock {\em Nature Rev. Phys.}, 2(1):10--12, 2019.

\bibitem{Schutz:1986gp}
B.~F. Schutz.
\newblock {Determining the Hubble Constant from Gravitational Wave
  Observations}.
\newblock {\em Nature}, 323:310--311, 1986.

\bibitem{Schutz:2001re}
B.~F. Schutz.
\newblock {Lighthouses of gravitational wave astronomy}.
\newblock In {\em {Proceedings, MPA / ESO / MPE / USM Conference on Lighthouses
  of the Universe: The Most Luminous Celestial Objects and their use for
  Cosmology: Garching, Munich, Germany, August 6-9, 2001}}, 2001.

\bibitem{Holz:2005df}
D.~E. Holz and S.~A. Hughes.
\newblock {Using gravitational-wave standard sirens}.
\newblock {\em Astrophys. J.}, 629:15--22, 2005.

\bibitem{Abbott:2017xzu}
B.~P. Abbott et~al.
\newblock {A gravitational-wave standard siren measurement of the Hubble
  constant}.
\newblock {\em Nature}, 551(7678):85--88, 2017.

\bibitem{Fishbach:2018gjp}
M.~Fishbach et~al.
\newblock {A Standard Siren Measurement of the Hubble Constant from GW170817
  without the Electromagnetic Counterpart}.
\newblock {\em Astrophys. J.}, 871(1):L13, 2019.
\newblock arXiv:1807.05667.

\bibitem{Soares-Santos:2019irc}
M.~Soares-Santos et~al.
\newblock {First Measurement of the Hubble Constant from a Dark Standard Siren
  using the Dark Energy Survey Galaxies and the LIGO/Virgo Binary–Black-hole
  Merger GW170814}.
\newblock {\em Astrophys. J.}, 876(1):L7, 2019.
\newblock arXiv:1901.01540.

\bibitem{Abbott:2019yzh}
B.~P. Abbott et~al.
\newblock {A gravitational-wave measurement of the Hubble constant following
  the second observing run of Advanced LIGO and Virgo}.
\newblock 2019.
\newblock arXiv:1908.06060.

\bibitem{TheLIGOScientific:2017qsa}
B.~P. Abbott et~al.
\newblock {GW170817: Observation of Gravitational Waves from a Binary Neutron
  Star Inspiral}.
\newblock {\em Phys. Rev. Lett.}, 119(16):161101, 2017.
\newblock arXiv:1710.05832.

\bibitem{2017Natur.551...85A}
B.~P. {Abbott}, R.~{Abbott}, T.~D. {Abbott}, et~al.
\newblock {A gravitational-wave standard siren measurement of the Hubble
  constant}.
\newblock {\em Nat}, 551:85--88, November 2017.

\bibitem{2019PhRvX...9a1001A}
B.~P. {Abbott}, R.~{Abbott}, T.~D. {Abbott}, et~al.
\newblock {Properties of the Binary Neutron Star Merger GW170817}.
\newblock {\em Phys.~Rev.~X}, 9(1):011001, Jan 2019.

\bibitem{LVCO2Cosmo}
B.~P. {Abbott}, R.~{Abbott}, T.~D. {Abbott}, et~al.
\newblock A gravitational-wave measurement of the hubble constant following the
  second observing run of advanced ligo and virgo.
\newblock page arXiv:\href{https://arxiv.org/abs/1908.06060}{1908.06060},
  August 2019.

\bibitem{DelPozzo:2011yh}
W.~Del~Pozzo.
\newblock {Inference of the cosmological parameters from gravitational waves:
  application to second generation interferometers}.
\newblock {\em Phys. Rev.}, D86:043011, 2012.

\bibitem{Gray:2019ksv}
R.~Gray et~al.
\newblock {Cosmological Inference using Gravitational Wave Standard Sirens: A
  Mock Data Challenge}.
\newblock 2019.

\bibitem{Reitze:2019iox}
D.~Reitze et~al.
\newblock {Cosmic Explorer: The U.S. Contribution to Gravitational-Wave
  Astronomy beyond LIGO}.
\newblock 2019.

\bibitem{Sathyaprakash:2012jk}
B.~Sathyaprakash et~al.
\newblock {Scientific Objectives of Einstein Telescope}.
\newblock {\em Class. Quant. Grav.}, 29:124013, 2012.
\newblock [Erratum: Class. Quant. Grav.30,079501(2013)].

\bibitem{Audley:2017drz}
P.~Amaro-Seoane et~al.
\newblock {Laser Interferometer Space Antenna}.
\newblock 2017.
\newblock arXiv:1702.00786.

\bibitem{Tamanini:2016zlh}
N.~Tamanini et al.
\newblock {Science with the space-based interferometer eLISA. III: Probing the
  expansion of the Universe using gravitational wave standard sirens}.
\newblock {\em JCAP}, 1604(04):002, 2016.

\bibitem{Tamanini:2016uin}
N.~Tamanini.
\newblock {Late time cosmology with LISA: probing the cosmic expansion with
  massive black hole binary mergers as standard sirens}.
\newblock {\em J. Phys. Conf. Ser.}, 840(1):012029, 2017.

\bibitem{Merloni:2012uf}
A.~Merloni et~al.
\newblock {eROSITA Science Book: Mapping the Structure of the Energetic
  Universe}.
\newblock 9 2012.

\bibitem{refId0}
{Kolodzig, Alexander} et al.
\newblock Agn and qsos in the erosita all-sky survey - i. statistical
  properties.
\newblock {\em A\&A}, 558:A89, 2013.

\bibitem{Lusso:2020fax}
E.~Lusso.
\newblock {Cosmology with quasars: predictions for eROSITA from a quasar Hubble
  diagram}.
\newblock 2 2020.

\bibitem{Barnett:2019rtg}
R.~Barnett et~al.
\newblock {Euclid preparation - V. Predicted yield of redshift 7 \ensuremath{<}
  z \ensuremath{<} 9 quasars from the wide survey}.
\newblock {\em Astron. Astrophys.}, 631:A85, 2019.

\bibitem{code_repo}
\url{https://github.com/lorenzsp/StandardSirensVSQuasars}.

\bibitem{1982ApJ...262L..17A}
Y.~{Avni} and H.~{Tananbaum}.
\newblock {On the cosmological evolution of the X-ray emission from quasars}.
\newblock {\em \apjl}, 262:L17--L21, November 1982.

\bibitem{Betoule:2014frx}
M.~Betoule et~al.
\newblock {Improved cosmological constraints from a joint analysis of the
  SDSS-II and SNLS supernova samples}.
\newblock {\em Astron. Astrophys.}, 568:A22, 2014.

\bibitem{banerjee2020cosmography}
A.~Banerjee et al.
\newblock {On cosmography in the cosmic dark ages: are we still in the dark?}
\newblock 9 2020.

\bibitem{emcee}
D.~{Foreman-Mackey}, D.~W. {Hogg}, D.~{Lang}, and J.~{Goodman}.
\newblock emcee: The mcmc hammer.
\newblock {\em PASP}, 125:306--312, 2013.

\bibitem{2019ApJ...873L..12S}
E.~{Sonbas}, K.~S. {Dhuga}, and E.~{G{\"o}{\u{g}}{\"u}{\textcommabelow s}}.
\newblock {Evidence of an X-Ray-Ultraviolet Spectral Correlation in
  Ultraluminous X-Ray Sources}.
\newblock {\em \apjl}, 873(2):L12, March 2019.

\bibitem{Salvestrini:2019thn}
F.~Salvestrini et al.
\newblock {Quasars as standard candles II: The non linear relation between UV
  and X-ray emission at high redshifts}.
\newblock {\em Astron. Astrophys.}, 631:A120, 2019.

\bibitem{Breivik:2017jip}
K.~Breivik et al.
\newblock {Characterizing Accreting Double White Dwarf Binaries with the Laser
  Interferometer Space Antenna and Gaia}.
\newblock {\em Astrophys. J.}, 854(1):L1, 2018.

\bibitem{Korol:2017qcx}
V.~Korol et al.
\newblock {Prospects for detection of detached double white dwarf binaries with
  Gaia, LSST and LISA}.
\newblock {\em Mon. Not. Roy. Astron. Soc.}, 470(2):1894--1910, 2017.

\bibitem{Korol:2018ulo}
V.~Korol, O.~Koop, and E.~M. Rossi.
\newblock {Detectability of double white dwarfs in the Local Group with LISA}.
\newblock {\em Astrophys. J.}, 866(2):L20, 2018.

\bibitem{Lau:2019wzw}
M.~Y.~M. Lau et al.
\newblock {Detecting Double Neutron Stars with LISA}.
\newblock {\em Mon. Not. Roy. Astron. Soc.}, 492(3):3061--3072, 2020.

\bibitem{Sesana:2016ljz}
A.~Sesana.
\newblock {Prospects for Multiband Gravitational-Wave Astronomy after
  GW150914}.
\newblock {\em Phys. Rev. Lett.}, 116(23):231102, 2016.

\bibitem{Babak:2017tow}
S.~Babak et al.
\newblock {Science with the space-based interferometer LISA. V: Extreme
  mass-ratio inspirals}.
\newblock {\em Phys. Rev.}, D95(10):103012, 2017.

\bibitem{Klein:2015hvg}
A.~Klein et~al.
\newblock {Science with the space-based interferometer eLISA: Supermassive
  black hole binaries}.
\newblock {\em Phys. Rev.}, D93(2):024003, 2016.

\bibitem{Caprini:2015zlo}
C.~Caprini et~al.
\newblock {Science with the space-based interferometer eLISA. II: Gravitational
  waves from cosmological phase transitions}.
\newblock {\em JCAP}, 1604(04):001, 2016.

\bibitem{Bartolo:2016ami}
N.~Bartolo et~al.
\newblock {Science with the space-based interferometer LISA. IV: Probing
  inflation with gravitational waves}.
\newblock {\em JCAP}, 1612:026, 2016.

\bibitem{Caprini:2018mtu}
C.~Caprini and D.~G. Figueroa.
\newblock {Cosmological Backgrounds of Gravitational Waves}.
\newblock {\em Class. Quant. Grav.}, 35(16):163001, 2018.

\bibitem{Eracleous:2019bal}
M.~Eracleous et al.
\newblock {An Arena for Multi-Messenger Astrophysics: Inspiral and Tidal
  Disruption of White Dwarfs by Massive Black Holes}.
\newblock 2 2019.

\bibitem{Wang:2019bbk}
Y.~Wang, F.~Wang, Y.~Zou, and Z.~Dai.
\newblock {A bright electromagnetic counterpart to extreme mass ratio
  inspirals}.
\newblock {\em Astrophys. J. Lett.}, 886(1):L22, 2019.

\bibitem{Zhang:2019dpy}
B.~Zhang.
\newblock {Charged Compact Binary Coalescence Signal and Electromagnetic
  Counterpart of Plunging Black Hole\textendash{}Neutron Star Mergers}.
\newblock {\em Astrophys. J.}, 873(2):L9, 2019.

\bibitem{Kyutoku:2016zxn}
K.~Kyutoku and N.~Seto.
\newblock {Gravitational-wave cosmography with LISA and the Hubble tension}.
\newblock {\em Phys. Rev.}, D95(8):083525, 2017.

\bibitem{DelPozzo:2017kme}
W.~Del~Pozzo, A.~Sesana, and A.~Klein.
\newblock {Stellar binary black holes in the LISA band: a new class of standard
  sirens}.
\newblock {\em Mon. Not. Roy. Astron. Soc.}, 475(3):3485--3492, 2018.

\bibitem{MacLeod:2007jd}
C.~L. MacLeod and C.~J. Hogan.
\newblock {Precision of Hubble constant derived using black hole binary
  absolute distances and statistical redshift information}.
\newblock {\em Phys. Rev.}, D77:043512, 2008.

\bibitem{laghi2020}
D.~Laghi et~al.
\newblock 2020.
\newblock \textit{in preparation}.

\bibitem{Palenzuela:2010nf}
C.~Palenzuela, L.~Lehner, and S.~L. Liebling.
\newblock {Dual Jets from Binary Black Holes}.
\newblock {\em Science}, 329:927, 2010.

\bibitem{2012AdAst2012E...3D}
M.~{Dotti}, A.~{Sesana}, and R.~{Decarli}.
\newblock {Massive Black Hole Binaries: Dynamical Evolution and Observational
  Signatures}.
\newblock {\em Advances in Astronomy}, 2012:940568, January 2012.

\bibitem{Giacomazzo:2012iv}
B.~Giacomazzo et al.
\newblock {General Relativistic Simulations of Magnetized Plasmas around
  Merging Supermassive Black Holes}.
\newblock {\em Astrophys. J.}, 752:L15, 2012.

\bibitem{Kocsis:2007yu}
B.~Kocsis, Z.~Haiman, and K.~Menou.
\newblock {Pre-Merger Localization of Gravitational-Wave Standard Sirens With
  LISA: Triggered Search for an Electromagnetic Counterpart}.
\newblock {\em Astrophys. J.}, 684:870--888, 2008.

\bibitem{OShaughnessy:2011nwl}
R.~O'Shaughnessy, D.~L. Kaplan, A.~Sesana, and A.~Kamble.
\newblock {Blindly detecting orbital modulations of jets from merging
  supermassive black holes}.
\newblock {\em Astrophys. J.}, 743:136, 2011.

\bibitem{2011ApJ...734L..37K}
D.~L. {Kaplan}, R.~{O'Shaughnessy}, A.~{Sesana}, and M.~{Volonteri}.
\newblock {Blindly Detecting Merging Supermassive Black Holes with Radio
  Surveys}.
\newblock {\em \apjl}, 734(2):L37, June 2011.

\bibitem{Haiman:2017szj}
Z.~Haiman.
\newblock {Electromagnetic chirp of a compact binary black hole: A phase
  template for the gravitational wave inspiral}.
\newblock {\em Phys. Rev.}, D96(2):023004, 2017.

\bibitem{Caprini:2016qxs}
C.~Caprini and N.~Tamanini.
\newblock {Constraining early and interacting dark energy with gravitational
  wave standard sirens: the potential of the eLISA mission}.
\newblock {\em JCAP}, 10:006, 2016.

\bibitem{Cai:2017yww}
R.-G. Cai, N.~Tamanini, and T.~Yang.
\newblock {Reconstructing the dark sector interaction with LISA}.
\newblock {\em JCAP}, 05:031, 2017.

\bibitem{Belgacem:2019pkk}
E.~Belgacem et~al.
\newblock {Testing modified gravity at cosmological distances with LISA
  standard sirens}.
\newblock {\em JCAP}, 07:024, 2019.

\bibitem{Hirata:2010ba}
C.~M. Hirata, D.~E. Holz, and C.~Cutler.
\newblock {Reducing the weak lensing noise for the gravitational wave Hubble
  diagram using the non-Gaussianity of the magnification distribution}.
\newblock {\em Phys.\ Rev.\ D}, 81:124046, 2010.

\bibitem{Kocsis:2005vv}
B.~Kocsis, Z.~Frei, Z.~Haiman, and K.~Menou.
\newblock {Finding the electromagnetic counterparts of cosmological standard
  sirens}.
\newblock {\em Astrophys.\ J.}, 637:27--37, 2006.

\bibitem{LiPhDThesis}
T.~G.~F. Li.
\newblock {Extracting Physics from Gravitational Waves}.
\newblock {\em Springer Thesis}, 2015.

\bibitem{NewMBHBCats}
LISA-Science-Group.
\newblock {\em in preparation}, 2021.

\bibitem{kassBayesFactors1995}
R.~E. Kass and A.~E. Raftery.
\newblock Bayes {{Factors}}.
\newblock {\em Journal of the American Statistical Association},
  90(430):773--795.

\bibitem{Risaliti_2015}
G.~Risaliti and E.~Lusso.
\newblock A hubble diagram for quasars.
\newblock {\em The Astrophysical Journal}, 815(1):33, Dec 2015.

\bibitem{Bull:2020cpe}
P.~Bull, M.~White, and A.~Slosar.
\newblock {Searching for dark energy in the matter-dominated era}.
\newblock 7 2020.

\bibitem{Bernal:2019gfq}
J.~L. Bernal, P.~C. Breysse, and E.~D. Kovetz.
\newblock {Cosmic Expansion History from Line-Intensity Mapping}.
\newblock {\em Phys. Rev. Lett.}, 123(25):251301, 2019.

\bibitem{Karkare:2018sar}
K.~S. Karkare and S.~Bird.
\newblock {Constraining the Expansion History and Early Dark Energy with Line
  Intensity Mapping}.
\newblock {\em Phys. Rev. D}, 98(4):043529, 2018.

\bibitem{Kovetz:2017agg}
E.~D. Kovetz et~al.
\newblock {Line-Intensity Mapping: 2017 Status Report}.
\newblock 9 2017.

\bibitem{Cusin:2020ezb}
G.~Cusin and N.~Tamanini.
\newblock {Characterisation of lensing selection effects for LISA massive black
  hole binary mergers}.
\newblock 11 2020.

\bibitem{Shapiro:2009sr}
C.~Shapiro, D.~Bacon, M.~Hendry, and B.~Hoyle.
\newblock {Delensing Gravitational Wave Standard Sirens with Shear and Flexion
  Maps}.
\newblock {\em Mon. Not. Roy. Astron. Soc.}, 404:858--866, 2010.

\bibitem{Marsat:2020rtl}
S.~Marsat, J.~G. Baker, and T.~Dal~Canton.
\newblock {Exploring the Bayesian parameter estimation of binary black holes
  with LISA}.
\newblock 2 2020.

\bibitem{Dahlen:2013fea}
T.~Dahlen et~al.
\newblock {A Critical Assessment of Photometric Redshift Methods: A CANDELS
  Investigation}.
\newblock {\em Astrophys. J.}, 775:93, 2013.

\bibitem{Ilbert:2013bf}
O.~Ilbert et~al.
\newblock {Mass assembly in quiescent and star-forming galaxies since z=4 from
  UltraVISTA}.
\newblock {\em Astron. Astrophys.}, 556:A55, 2013.

\bibitem{Kruhler:2010jw}
T.~Kruhler et~al.
\newblock {Photometric redshifts for GRB afterglows from GROND and Swift/UVOT}.
\newblock {\em Astron. Astrophys.}, 526:A153, 2011.

\bibitem{Dawson:2015wdb}
K.~S. Dawson et~al.
\newblock {The SDSS-IV extended Baryon Oscillation Spectroscopic Survey:
  Overview and Early Data}.
\newblock {\em Astron. J.}, 151:44, 2016.

\bibitem{Amendola:2016saw}
L.~Amendola et~al.
\newblock {Cosmology and fundamental physics with the Euclid satellite}.
\newblock {\em Living Rev. Rel.}, 21(1):2, 2018.

\end{thebibliography}

\end{document}